\documentclass[aps,prd,twocolumn,superscriptaddress,nofootinbib,longbibliography,preprintnumbers,floatfix]{revtex4-2}

\usepackage[normalem]{ulem}
\usepackage{bm}
\usepackage[utf8]{inputenc}
\usepackage{amsmath}
\usepackage{amssymb}
\usepackage{multirow}
\usepackage{amsfonts}
\usepackage{upgreek}
\usepackage{subeqnarray}
\usepackage{graphicx}
\usepackage[dvipsnames]{xcolor}
\usepackage{yfonts}
\usepackage{siunitx}
\usepackage{comment}
\usepackage[T1]{fontenc}

\usepackage{epsfig}
\usepackage[colorlinks]{hyperref}
\usepackage{xspace}

\usepackage{tikz}
\usetikzlibrary{tikzmark}
\usetikzlibrary{positioning}

\newcommand{\stkout}[1]{\ifmmode\text{\sout{\ensuremath{#1}}}\else\sout{#1}\fi}

\definecolor{lime}{HTML}{A6CE39}
\DeclareRobustCommand{\orcidicon}{\hspace{-1mm}
	\begin{tikzpicture}
	\draw[lime, fill=lime] (0,0) 
	circle [radius=0.16] 
	node[white] {{\fontfamily{qag}\selectfont \tiny \,ID}};
	\draw[white, fill=white] (-0.0525,0.095) 
	circle [radius=0.007];
	\end{tikzpicture}
	\hspace{-3mm}
}

\foreach \x in {A, ..., Z}{\expandafter\xdef\csname orcid\x\endcsname{\noexpand\href{https://orcid.org/\csname orcidauthor\x\endcsname}
			{\noexpand\orcidicon}}
}

\hypersetup{linkcolor=BrickRed,citecolor=Green,
filecolor=Mulberry,
urlcolor=NavyBlue,
menucolor=BrickRed,
runcolor=Mulberry
}

\newcommand{\appsection}[1]{\section{\MakeUppercase{#1}}}

\newcommand{\vrho}{\varrho}

\newcommand{\dd}{\mathrm{d}}
\newcommand{\had}{\hat{a}^\dagger}
\newcommand{\ha}{\hat{a}}
\newcommand{\hbd}{\hat{b}^\dagger}
\newcommand{\hb}{\hat{b}}
\renewcommand{\i}{\mathrm{i}}
\newcommand{\ux}{\vec{u}_x}
\newcommand{\uy}{\vec{u}_y}
\newcommand{\uz}{\vec{u}_z}
\newcommand{\Id}{\mathbb{I}}

\begin{document}
\preprint{INT-PUB-26-025, N3AS-26-010, RIKEN-iTHEMS-Report-26}

\title{Collective neutrino oscillations: Many-body non-forward effects and non-classicality}

\author{Julien Froustey\orcidA{}}
\email[]{julien.froustey@ific.uv.es}
\affiliation{Institut de Física Corpuscular (IFIC), CSIC-Universitat de València, Parc Científic UV, C/ Catedrático José Beltrán 2, E-46980 Paterna (València), Spain}

\author{Ermal Rrapaj\orcidB{}}
\email[]{ermalrrapaj@lbl.gov}
\affiliation{Lawrence Berkeley National Laboratory, Berkeley, California 94720, USA}
\affiliation{Department of Physics, University of California Berkeley, Berkeley, California 94720, USA}
\affiliation{RIKEN iTHEMS, Wako, Saitama 351-0198, Japan}

\author{Yuhao Liu\orcidC{}}
\affiliation{Department of Computer and Information Science, University of Pennsylvania, Philadelphia, Pennsylvania 19104, USA}

\author{Gushu Li\orcidD{}}
\affiliation{Department of Computer and Information Science, University of Pennsylvania, Philadelphia, Pennsylvania 19104, USA}

\author{Costin Iancu\orcidE{}}
\affiliation{Lawrence Berkeley National Laboratory, Berkeley, California 94720, USA}

\author{Vincenzo Cirigliano\orcidF{}{$\ $}}
\affiliation{Institute for Nuclear Theory, University of Washington, Seattle, WA 91195-1550, USA}

\begin{abstract}
Neutrino evolution in dense astrophysical environments is typically described either within a quantum kinetic framework, which neglects the build-up of multi-body correlations, or through simplified many-body calculations that allow significant entanglement to develop. In this work, we compare these two approaches in a simple neutrino-gas configuration, with particular emphasis on the role of non-forward scattering processes. These effects are incorporated either through a collision term in the kinetic description, or by considering the full neutrino–neutrino many-body Hamiltonian. We highlight differences between the two descriptions in both their characteristic timescales and asymptotic behavior. Motivated by the natural suitability of quantum computing for many-body calculations, we further investigate the non-classicality of neutrino evolution, discussing Trotter error scaling, along with the associated costs of constructing quantum circuits in terms of entangling gates and non-Clifford gates. We find that the resources needed for neutrino many-body evolution are on the low end of typical high-energy physics problems and on the mid to high end with respect to quantum chemistry problems. For the full Hamiltonian, resource requirements increase relative to the truncated version. We emphasize the importance of efficient fermion-to-qubit encodings, which are essential for reducing the substantial computational resources required for such simulations.
\end{abstract}

\maketitle

\section{Introduction}

Neutrinos play a key role in dense astrophysical environments such as core-collapse supernovae (CCSNe) and neutron star mergers (NSMs), influencing notably their dynamics or the synthesis of heavy elements~\cite{Janka:2012wk,Mezzacappa:2020oyq,Burrows:2020qrp,Foucart:2022bth,Fischer:2023ebq,Wang:2023tso}. The typical temperatures in these media (dozens of MeV) lead to a hierarchy between the populations of charged leptons, with mostly negligible amounts of muon and tau leptons. As a consequence, different neutrino flavors experience different interactions, with a specific role played by electron neutrinos and antineutrinos compared to the other species. This suggests that \emph{flavor transformation} could be a crucial physical mechanism in CCSNe and NSMs, affecting for instance the energy deposition leading to a successful supernova explosion, or the neutron-to-proton ratios and eventual nucleosynthesis yields.

The traditional framework used to model neutrino transport while accounting for flavor oscillations relies on the quantum kinetic equation (QKE) approach~\cite{SiglRaffelt,Vlasenko:2013fja,Blaschke:2016xxt,Volpe:2013uxl,Volpe:2015rla,Froustey:2020mcq}. It reduces the \emph{a priori} many-body entangled problem describing the evolution of neutrinos to a set of one-body equations, and is thus commonly referred to as a ``mean-field'' description. A molecular chaos approximation allows one to derive a momentum-exchanging collision term, thus fully generalizing the classical Boltzmann equation approach.

Mean-field studies have revealed a rich variety of flavor conversion mechanisms, such as MSW resonances, matter-neutrino resonances, or so-called flavor instabilities (see the reviews~\cite{Duan:2010bg,Tamborra:2020cul,Capozzi:2022slf,Volpe:2023met,Johns:2025mlm}). Since it neglects many-body quantum correlations, the adequacy of this approach has been questioned over the years, see~\cite{Pantaleone:1992eq,Bell:2003mg,Friedland:2003dv,Friedland:2003eh,Friedland:2006ke} for early works and \cite{Patwardhan:2022mxg,Balantekin:2023qvm} for recent reviews. A number of studies have identified situations in which substantial entanglement growth can occur, with a rich phenomenology including spectral splits~\cite{Cervia:2019res,Patwardhan:2021rej,Hite:2026fsj}, multi-angle~\cite{Rrapaj:2019pxz,Roggero:2022hpy} and three-flavor~\cite{Siwach:2024jet} effects, or quantum equilibration~\cite{Martin:2023gbo}. The associated changes in neutrino distributions can affect the evolution of dense astrophysical systems, a possibility explored in large-scale simulations of NSMs~\cite{Qiu:2025kgy,Qiu:2025ybw}, and for nucleosynthesis in CCSN environments~\cite{Balantekin:2023ayx}.

The exponential growth of the size of the Hilbert space with the number of neutrinos in the system strongly limits the capability of classical computers to describe a many-body neutrino system, with the exception of some highly symmetric geometries~\cite{Xiong:2021evk,Roggero:2022hpy,Martin:2021bri}. Various techniques have been developed to tackle this difficulty, such as tensor network representations~\cite{Roggero:2021asb,Roggero:2021fyo,Cervia:2022pro}, the use of a Bethe ansatz~\cite{Pehlivan:2011hp,Birol:2018qhx,Patwardhan:2019zta}, or phase-space methods~\cite{Lacroix:2022krq,Lacroix:2024pbb}. This problem is also, for the same reasons, a natural playground for calculations on quantum computers, with several recent studies using different quantum hardware~\cite{Hall:2021rbv, Yeter-Aydeniz:2021olz, Jha:2021itm, Amitrano:2022yyn, Illa:2022zgu, Siwach:2023wzy, Turro:2024shh, Singh:2024vpu, Spagnoli:2025etu, Tripathi:2025cok, Bleau:2026iiq, Heimsoth:2026ozm}. The ``non-classicality'' of neutrino evolution, determined by metrics which quantify the need for an actual quantum computation, was notably studied in~\cite{Chernyshev:2024pqy}.

All the previously mentioned studies focusing on the many-body neutrino problem do not use the full Hamiltonian describing neutrino-neutrino interactions, but a ``truncated'' expression which only retains the so-called forward and exchange terms (for which the pairs of incoming and outgoing momenta of the interacting neutrinos are identical). These terms are indeed the only relevant ones in the mean-field limit, but singling them out in a many-body approach is not \emph{a priori} justified~\cite{Johns:2023ewj}. The first study of many-body neutrino oscillations with the full Hamiltonian was performed in~\cite{Cirigliano:2024pnm}, and notably showed that the timescales of flavor and kinetic equilibration were similar, at odds with the quantum kinetic picture. More recently, a semi-classical treatment including non-forward scattering was developed in~\cite{Carlson:2026mir}, and an approach with wave packets instead of plane waves for a highly symmetric problem was used in~\cite{Cervia:2025pfg}. It was also recently shown that non-forward terms could significantly enhance neutrino \emph{helicity} transformation via many-body effects~\cite{Xu:2026mlq}. 

In this paper, we extend the work of~\cite{Cirigliano:2024pnm} in two directions. First, since a feature of the full Hamiltonian is to allow for non-forward momentum exchange, which is impossible in the mean-field approximation, we make a more adequate comparison by implementing a QKE collision term for a discrete set of momenta. We find that, as expected, the dynamical timescales are very different between the QKE and many-body calculations, but the asymptotic state is also different, a feature of the inclusion (or not) of multi-body correlations. Second, we confirm that the non-forward terms in the Hamiltonian significantly change the evolution of one-body observables, meaning that, in a many-body framework, using the truncated Hamiltonian cannot be justified. We follow through in estimating the associated cost for a quantum simulation of this problem, determining the scaling of the Trotter error with the number of degrees of freedom, and numerically obtaining the scaling for the number of quantum gates required. This latter point depends on the specific fermion-to-qubit encoding adopted~\cite{Liu:2024rtx}.

This paper is organized as follows. In Sec.~\ref{sec:evolution_equations}, we introduce the Hamiltonian describing the neutrino system, with conventions and details in Appendix~\ref{app:deriv_H}. We then present the discrete form of the QKE adapted to our momentum grid, with further details in Appendix~\ref{app:QKE}. In Sec.~\ref{sec:time_evolution}, we study the evolution of one-body observables for an initial configuration of 3 neutrinos of different flavors, comparing the mean-field, QKE, truncated and full many-body calculations. A simpler, two-beam neutrino example is shown in Appendix~\ref{app:2beams}. We investigate the deviation from classicality, and especially the changes incurred from using the full instead of the truncated expression of the many-body Hamiltonian, in Sec.~\ref{sec:quantumness}. Details of the Trotter error calculations are gathered in Appendix~\ref{app:Trotter_details}. We summarize our findings in Sec.~\ref{sec:conclusion}.

Throughout this work, we use natural units for which $\hbar = c = k_B = 1$.

\medskip

\section{Neutrino evolution equations}
\label{sec:evolution_equations}

We first introduce the formalisms used to describe the dynamics of an ensemble of neutrinos, beginning with a full many-body treatment and followed by a quantum kinetic approach that reduces the system to one-body observables.

\subsection{Neutrino Hamiltonian}

In the system we consider, each one-particle state can be described by its momentum ($\vec{p}$), flavor ($\alpha = e, \mu, \tau$), and helicity. For the latter, we will consider ultrarelativistic neutrinos such that only the left-handed helicity state contributes. We can thus define the annihilation and creation fermionic operators $\ha_\alpha(\vec{p}),\had_\alpha(\vec{p})$, which satisfy the anticommutation relation
\begin{equation}
\label{eq:had_ha}
    \left\{ \had_{\alpha}(\vec{p}),\ha_\beta(\vec{q})\right\} = (2 \pi)^3 \delta^{(3)}(\vec{p}-\vec{q}) \, \delta_{\alpha \beta} \, .
\end{equation}

The Hamiltonian $H = H_\mathrm{vac} + H_{\nu \nu}$ driving the dynamics of the system is composed of a one-body part, associated to vacuum oscillations ($H_\mathrm{vac}$, “vacuum Hamiltonian”), and a two-body part describing neutrino-neutrino self-interactions ($H_\mathrm{\nu \nu}$, “self-interaction Hamiltonian”). We do not consider the presence of non-neutrino species such as charged leptons, which would add another two-body term to the Hamiltonian. Using the Standard Model of weak interactions in the low energy limit (Fermi theory), the total Hamiltonian is given by 

\begin{widetext}
\begin{multline}
\label{eq:H_full}
    H = \sum_{\alpha, \beta} \int{\frac{\dd^3 \vec{p}}{(2 \pi)^3} \omega^{\alpha}_{\beta}(p) \, \had_\alpha(\vec{p}) \ha_\beta(\vec{p})}  + \sqrt{2} G_F \sum_{\alpha,\beta} \int \frac{\dd^3 \vec{p}_1}{(2 \pi)^3 \sqrt{2 p_1}} \frac{\dd^3 \vec{p}_2}{(2 \pi)^3 \sqrt{2 p_2}} \frac{\dd^3 \vec{p}_3}{(2 \pi)^3 \sqrt{2 p_3}} \frac{\dd^3 \vec{p}_4}{(2 \pi)^3 \sqrt{2 p_4}} \\ \times (2 \pi)^3 \, \delta^{(3)}(\vec{p}_1 + \vec{p}_2 - \vec{p}_3 - \vec{p}_4)
    \times \sqrt{p_1p_2p_3p_4} \, \mathcal{V}(\vec{p}_1,\vec{p}_2,\vec{p}_3,\vec{p}_4) \, \had_{\alpha}(\vec{p}_1) 
    \ha_{\alpha}(\vec{p}_3)\had_{\beta}(\vec{p}_2) \ha_{\beta}(\vec{p}_4) \, ,
\end{multline}
where the interaction term is,
\begin{equation}
\label{eq:v_nunu_general}
    \mathcal{V}(\vec{p}_1,\vec{p}_2,\vec{p}_3,\vec{p}_4) \equiv  \frac{\left[ (1 - \hat{p}_{1,z})(\hat{p}_{2,x}-\i \hat{p}_{2,y}) - (\hat{p}_{1,x} - \i \hat{p}_{1,y})(1 - \hat{p}_{2,z})\right] \left[ (1 - \hat{p}_{3,z})(\hat{p}_{4,x}+\i \hat{p}_{4,y}) - (\hat{p}_{3,x} + \i \hat{p}_{3,y})(1 - \hat{p}_{4,z})\right]}{\sqrt{(1 - \hat{p}_{1,z})(1-\hat{p}_{2,z})(1-\hat{p}_{3,z})(1-\hat{p}_{4,z})}} \, ,
\end{equation}
with the unit vector denoted as $\hat{\vec{p}}_i \equiv \vec{p}_i/\lvert \vec{p}_i\rvert$. Conventions and details are gathered in Appendix~\ref{app:deriv_H}.
\end{widetext}
This expression agrees with Eq.~(17) in \cite{Cirigliano:2024pnm}, up to an overall phase difference which has no physical consequence (see Appendix~\ref{app:deriv_H} for the origin of this phase difference). The vacuum Hamiltonian coefficients read
\begin{align}
    \omega^{a}_{b}(p) &= \sqrt{p^2 + m_a^2} \, \delta^{a}_{b} \\
\intertext{in the mass basis, and}
    \omega^{\alpha}_{\beta}(p) &= \sum_{a=1}^{3} \sqrt{p^2 + m_a^2} U_{\alpha a} U_{\beta a}^*
\end{align}
in the flavor basis, with $U$ the Pontecorvo-Maki-Nakagawa-Sakata (PMNS) matrix. In astrophysical environments, neutrinos have typical energies of the order of a few $\mathrm{MeV}$, such that $p \gg m_a$. We then have:
\begin{equation}
\label{eq:energy_ultrarel}
    \sqrt{p^2 + m_a^2} \simeq p + \frac{m_a^2}{2 p} = p + \frac{m_1^2}{2p} + \frac{\Delta m_{a1}^2}{2p} \, ,
\end{equation}
where the mass-squared differences $\Delta m_{ab}^2 \equiv m_a^2 - m_b^2$ are measured in neutrino oscillation experiments~\cite{ParticleDataGroup:2024cfk}. Neglecting $m_1^2 \ll p^2$, we thus have:
\begin{multline}
    H_\mathrm{vac} = \int{\frac{\dd^3 \vec{p}}{(2 \pi)^3}} \Bigg[\sum_{\alpha} p \, \had_\alpha(\vec{p}) \ha_\alpha(\vec{p}) \\
    + \sum_{\alpha, \beta} \sum_{a=1}^{3} \delta \omega_a U_{\alpha a} U_{\beta a}^*  \had_\alpha(\vec{p}) \ha_\beta(\vec{p}) \Bigg] \, ,
\end{multline}
with $\delta \omega_a \equiv \Delta m_{a1}^2/2p$. An alternative\footnote{This choice has no physical consequence, since the oscillation frequency is set by the differences $\delta \omega_a - \delta \omega_b$, which are in any case equal to $\Delta m_{ab}^2/2p$.} consists in separating the sum of neutrino masses squared in \eqref{eq:energy_ultrarel}, such that the vacuum Hamiltonian takes the same final form with $\delta \omega_a = \sum_{b} \Delta m_{ab}^2/6p$.

\subsection{Discretization of momenta}

In order to numerically implement the evolution of a system of neutrinos under the Hamiltonian~\eqref{eq:H_full}, we need a discrete grid of momenta. For three dimensions, we consider the momenta:
\begin{equation}
\label{eq:pgrid}
    \vec{p} = \frac{2 \pi}{L_x} n_x \ux + \frac{2 \pi}{L_y} n_y \uy + \frac{2 \pi}{L_z} n_z \uz \, ,
\end{equation}
where $(n_x,n_y,n_z) \in \mathbb{Z}^3$ (excluding the case $n_x=n_y=n_z=0$). The “quantization volume” is $V=L_x\times L_y \times L_z$. In the applications we will consider, we will restrict to $L_x = L_y = L_z$.

We can thus write a discretized version of the Hamiltonian~\eqref{eq:H_full}. Denoting the grid of momenta $\{\vec{p}_i, i=1,\dots,N\}$, we have the following replacement rule for continuous integrals:
\begin{equation}
\label{eq:discrete_sum}
    \int{\frac{\dd^3 \vec{p}}{(2 \pi)^3} f(\vec{p})} \longrightarrow \frac{1}{V} \sum_{i=1}^{N}{f(\vec{p}_i)} \, .
\end{equation}
Similarly, Dirac delta functions become Kronecker deltas with the correspondence:
\begin{equation}
\label{eq:discrete_delta}
    (2 \pi)^3 \delta^{(3)}(\vec{p}_{i} - \vec{p}_{j}) \longrightarrow V \delta_{i, j}
\end{equation}
Finally, we define the dimensionless creation/annihilation operators
\begin{equation}
    \ha_{\alpha,i} \equiv \frac{1}{\sqrt{V}} \, \ha_{\alpha}(\vec{p}_i) \, ,
\end{equation}
which satisfy the discrete version of the canonical anticommutation relation, $\{\had_{\alpha,i},\ha_{\beta,j}\} = \delta_{\alpha \beta} \delta_{ij}$ [as can be checked using Eqs.~\eqref{eq:had_ha} and \eqref{eq:discrete_delta}].

The discretized version of the Hamiltonian~\eqref{eq:H_full} is thus $H = H_\mathrm{vac} + H_{\nu \nu}$, with
\begin{equation}
\label{eq:H_full_discrete}
\begin{aligned}
   H_\mathrm{vac} &= 
    \sum_{\alpha,\beta} \sum_{i=1}^{N} \omega^{\alpha}_{\beta}(\vec{p}_i) \, \had_{\alpha,i} \ha_{\beta,i} \, ,  \\  
    H_{\nu \nu} &= \frac{G_F}{2 \sqrt{2} V} \sum_{\alpha,\beta} \sum_{i_1,i_2,i_3,i_4=1}^{N} h_{\alpha,\beta,i_1,i_2,i_3,i_4} \, , 
\end{aligned}
\end{equation}
where the coefficients of the two-body part read
\begin{multline}
   h_{\alpha,\beta,i_1,i_2,i_3,i_4} \equiv \mathcal{V}(\vec{p}_{i_1},\vec{p}_{i_2},\vec{p}_{i_3},\vec{p}_{i_4}) \, \had_{\alpha,i_1} \ha_{\alpha,i_3}\had_{\beta,i_2} \ha_{\beta,i_4} \\ \times \delta_{\vec{p}_{i_4} = \vec{p}_{i_1}+\vec{p}_{i_2}-\vec{p}_{i_3}} \, ,
\end{multline}
with $\delta_{\vec{p}_{i_4} = \vec{p}_{i_1}+\vec{p}_{i_2}-\vec{p}_{i_3}}$ a Kronecker-delta function for momentum conservation.

For realistic neutrino systems, with typical energies of order $\sim \mathrm{MeV}$, the kinetic energy contribution to $ \omega^{\alpha}_{\beta}(\vec{p})$ is orders of magnitude larger than the other terms. This has an interesting consequence, discussed in~\cite{Cirigliano:2024pnm}: any two-body interaction term $h_{\alpha,\beta,i_1,i_2,i_3,i_4}$ which does not satisfy kinetic energy conservation is dynamically irrelevant. Written in the basis of two-body states, such a term would correspond to an ``off-diagonal transition'' between blocks of different total kinetic energy. On the finite grid of momenta we will consider, any nonzero kinetic energy difference will also be typically of order $\sim \mathrm{MeV}$; therefore, the energy split between the two blocks will be orders of magnitude larger than the off-diagonal perturbation $h_{\alpha,\beta,i_1,i_2,i_3,i_4}$. As a consequence, these transitions are suppressed and \emph{pairwise kinetic energy is dynamically conserved}. This feature is studied in details in~\cite{Cirigliano:2024pnm}, and we use it by restricting $H_{\nu \nu}$ to momentum combinations which satisfy both momentum and kinetic energy conservation.

\subsubsection{Forward/exchange limit}

Based on the coherent enhancement of some terms in the QKE formalism at the mean-field level (see Sec.~\ref{subsec:QKE}), many studies have restricted the many-body Hamiltonian interaction part to a subset of processes, namely, the forward [$(\vec{p}_3,\vec{p}_4) = (\vec{p}_1,\vec{p}_2)$] and exchange [$(\vec{p}_3,\vec{p}_4)=(\vec{p}_2,\vec{p}_1)$] terms. In those cases, we have:
\begin{equation}
\label{eq:V_forward}
    \begin{aligned}
        \mathcal{V}(\vec{p}_1,\vec{p}_2,\vec{p}_1,\vec{p}_2) &= + 2 (1 - \hat{\vec{p}}_1 \cdot \hat{\vec{p}}_2) \, , \\
        \mathcal{V}(\vec{p}_1,\vec{p}_2,\vec{p}_2,\vec{p}_1) &= - 2 (1 - \hat{\vec{p}}_1 \cdot \hat{\vec{p}}_2) \, .
    \end{aligned}
\end{equation}
The forward/exchange part of the self-interaction Hamiltonian (or “truncated” Hamiltonian as in~\cite{Cirigliano:2024pnm}) is thus, after anticommuting some $\ha, \had$,
\begin{multline}
\label{eq:H_forward}
    H_{\nu \nu}^{\mathrm{(f/e)}} = \frac{G_F}{\sqrt{2}V} \sum_{\alpha, \beta} \sum_{i_1 \neq i_2}{(1- \hat{\vec{p}}_{i_1}\cdot \hat{\vec{p}}_{i_2}}) \\\left[\had_{\alpha,i_1} \ha_{\alpha,i_1} \had_{\beta,i_2}\ha_{\beta,i_2} - \had_{\alpha,i_1} \ha_{\alpha, i_2}\had_{\beta,i_2}\ha_{\beta,i_1}\right] \, .
\end{multline}
Written for two flavors, this expression agrees with, for instance, Eq.~(18) in~\cite{Balantekin:2006tg}, up to terms proportional to the identity in SU(2) space.

\subsubsection{Schrödinger equation}

Given the neutrino Hamiltonian~\eqref{eq:H_full_discrete}, one can solve the Schrödinger equation
\begin{equation}
    \label{eq:schrodi_full}
    \i \frac{\dd |\Psi \rangle}{\dd t} = H \, |\Psi \rangle \, ,
\end{equation}
with $|\Psi\rangle$ the $N$-body quantum state of the system. The solutions of Eq.~\eqref{eq:schrodi_full} will be referred to as involving the ``full Hamiltonian.''

For comparison purposes with the literature, we will also present results where the interaction Hamiltonian is (a priori) arbitrarily restricted to the forward/exchange terms~\eqref{eq:H_forward}. We will thus refer to as ``truncated Hamiltonian'' the case corresponding to the equation
\begin{equation}
    \label{eq:schrodi_fwd}
    \i \frac{\dd |\Psi \rangle}{\dd t} = \left(H_\mathrm{vac} + H_{\nu \nu}^{\mathrm{(f/e)}}\right) |\Psi \rangle \, .
\end{equation}
%

\subsection{Discrete Quantum Kinetic Equation}
\label{subsec:QKE}

Instead of solving directly the $N$-body Schrödinger equation, the quantum kinetic approach consists in deriving an evolution equation written at the level of one-body observables only. It is often framed in the literature as being equivalent to the mean-field approximation (which also reduces the system to a collection of one-body quantities), but its general expression includes a \emph{collision term} which describes momentum exchange. We refer to, e.g.,~\cite{SiglRaffelt,Volpe:2013uxl,Volpe:2015rla,Froustey:2020mcq,Vlasenko:2013fja,Blaschke:2016xxt}, for details on the derivations of the so-called quantum kinetic equation (QKE), which generalizes the classical Boltzmann equation to include the phenomenon of flavor mixing. In the following, we only introduce the necessary physical ingredients for the upcoming discussion, and we present the form of the QKE suited for a system with discrete momenta, with additional details given in Appendix~\ref{app:QKE}.

We define the one-body density matrix
\begin{equation}
    \label{eq:rho_one_body}
    \vrho^{\alpha,i}_{\beta,j} \equiv \langle \had_{\beta,j} \ha_{\alpha,i}\rangle \, ,
\end{equation}
where $\langle \cdots \rangle = \langle \Psi |\cdots | \Psi\rangle$, or more generally for a mixed state with the $N$-body density matrix $\hat{\rho}$, $\langle \cdots \rangle = \mathrm{Tr}(\hat{\rho} \cdots)$. Since the Hamiltonian~\eqref{eq:H_full_discrete} conserves momentum, and we consider initial states of well-defined total momentum, the only nonzero entries of $\varrho$ are momentum-diagonal:
\begin{equation}
\label{eq:def_densitymatrix}
    \vrho^{\alpha,i}_{\beta,j} \equiv \delta_{ij} \, (f_i)^{\alpha}_{\beta} \, .
\end{equation}
It is possible to rewrite the exact equation~\eqref{eq:schrodi_full} as a \emph{hierarchy} of equations written for the one-body density matrix~\eqref{eq:rho_one_body} and the higher-order equivalent quantities (two-body, three-body, etc). This approach, corresponding to the Bogoliubov–Born–Green–Kirkwood–Yvon (BBGKY) hierarchy, was first introduced in~\cite{Volpe:2013uxl} for a system of neutrinos. Obtaining a finite system of equations then requires to close the hierarchy. The simplest assumption is the \emph{mean-field} (or Hartree-Fock) approximation, where the two-body density matrix is assumed to be a simple product of one-body contributions,
\begin{multline}
\label{eq:HartreeFock}
    \langle \had_{\gamma,k} \had_{\beta,j} \ha_{\alpha,i} \ha_{\delta, l} \rangle \approx \langle \had_{\gamma,k} \ha_{\delta,l}\rangle \langle \had_{\beta,j} \ha_{\alpha,i}\rangle \\ - \langle \had_{\gamma,k} \ha_{\alpha,i}\rangle \langle \had_{\beta,j} \ha_{\delta,l}\rangle \, .
\end{multline}
In other words, we neglect two-body correlations in the system. In this approximation, the only parts of the Hamiltonian that give a nonzero contribution are the forward/exchange terms. The mean-field equation then reads~\cite{SiglRaffelt,Qian:1994wh,Volpe:2013uxl}
\begin{equation}
\label{eq:meanfield}
    \dot{f}_{i} = - \i \left[\omega^{\alpha}_{\beta}(p_{i}), f_{i}\right] - \i \frac{\sqrt{2} G_F}{V} \sum_{j}{(1 - \hat{\vec{p}}_i \cdot \hat{\vec{p}}_j) \left[f_{j},f_{i}\right]} \, .
\end{equation}
If one does not make the mean-field approximation and includes the leftover correlations in~\eqref{eq:HartreeFock}, the evolution equation~\eqref{eq:meanfield} now includes a term depending on the two-body correlations~\cite{Volpe:2013uxl,Froustey:2020mcq}. Since the goal of the QKE is to describe neutrino transport, its derivation follows the assumptions which underlie the Boltzmann transport approach. Specifically, one assumes that neutrinos experience collisions described by individual scatterings, with the \emph{molecular chaos} ansatz: before each scattering, the incoming particles are supposed to be uncorrelated. This allows one\footnote{For details on this subtle step, see for instance Sec.~2.4 in~\cite{Lacroix2004}, or Sec.~6.1 in~\cite{Fidler:2017pkg}.} to transform the additional term in~\eqref{eq:meanfield} into a \emph{collision term}, such that the QKE reads
\begin{equation}
\label{eq:QKE}
    \dot{f}_{i} = - \i \left[\omega^{\alpha}_{\beta}(p_{i}), f_{i}\right] - \i \frac{\sqrt{2} G_F}{V} \sum_{j}{(1 - \hat{\vec{p}}_i \cdot \hat{\vec{p}}_j) \left[f_{j},f_{i}\right]} + \mathcal{C}_i \, ,
\end{equation}
with $\mathcal{C}_i$ the collision term, given in Eq.~\eqref{eq:discrete_collision}. In the existing literature, many-body calculations have always been compared with the mean-field equation. In the following, we also implement the version of the QKE including collisions~\eqref{eq:QKE}, written for a discrete grid of momenta.

\section{Time evolution of the neutrino system}
\label{sec:time_evolution}

\subsection{Setup}

\begin{figure}[ht]
    \centering
    \includegraphics[width=0.65\linewidth]{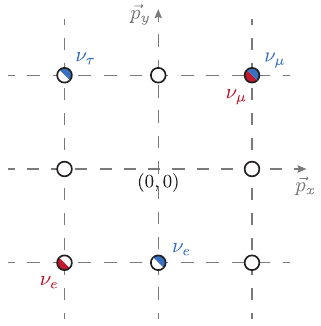}
    \caption{Two-dimensional grid of momenta used in this work. The states initially occupied in our test calculation are shown in blue, while the two-neutrino case shown in red is discussed in Appendix~\ref{app:2beams}. The momentum mode $\vec{p} = \vec{0}$ is excluded, as we restrict ourselves to ultrarelativistic neutrinos.}
    \label{fig:pgrid}
\end{figure}

In the following, and for computation reasons, we will restrict to two-dimensional systems with a low number of momentum modes. The grid of momenta and the initial configurations discussed in this work are shown in Fig.~\ref{fig:pgrid}. 

The parameters we consider are inspired by the conditions that can be found in a supernova, similarly to the approach of Ref.~\cite{Cirigliano:2024pnm}. Strictly speaking, there is only one free parameter in our model, the box size $L$, which sets the unit momentum $p_0 \equiv 2 \pi / L$, and the self-interaction strength $\mu \equiv G_F/(2 \sqrt{2} V)$. However, given our limited computational power, we cannot describe with a few dozen momentum bins the conditions that would correspond to both a typical density (important for the two-body interaction term) and a typical neutrino energy (important for the one-body term). As a consequence, we set independently $p_0$ and $V$, which allows us to obtain more ``physical'' timescales in our results.

Specifically, considering a temperature $T = 0.5 \, \mathrm{MeV}$, we want our simulation box to match an equilibrium density if there were a large number of neutrinos present, typically 100. As such, the volume $V$ is set by the equation
\begin{equation}
\label{eq:volume_simu}
    n = \frac{100}{V} = \frac{3 \zeta(3)}{4 \pi^2} T^3 \, ,
\end{equation}
which leads to a self-interaction strength
\begin{equation}
\label{eq:self_strength}
    \mu = \frac{G_F}{2 \sqrt{2} V} \simeq 4.7 \times 10^{-10} \ \mathrm{eV} \, ,
\end{equation}
corresponding to a timescale $\mu^{-1} \simeq 1.4 \, \mathrm{\mu s}$. The typical momentum of neutrinos, for this temperature, is $p_0 = 3T = 1.5 \, \mathrm{MeV}$, and the maximum momentum mode is $p_\mathrm{max}=\sqrt{2}p_0$. As a result, the inverse vacuum frequencies are in the range $[3.5 \mu^{-1}, 167 \mu^{-1}]$. In a more realistic setting, we would expect a stronger two-body coupling, associated with a nanosecond timescale (the typical scale of fast flavor oscillations~\cite{Richers_review}).\footnote{If we were to set $p_0$ to the unit momentum $2 \pi/L$ consistent with \eqref{eq:volume_simu}, it would correspond to an energy of $0.3 \, \mathrm{MeV}$, and our limited grid would only describe low-momentum modes, while we want to describe the ``typical'' evolution.} For computational purposes, we use a dimensionless time variable $\tilde{t} = \mu t$, or equivalently measure the time in units of $\mu^{-1}$.

We will denote $\lvert \nu_\alpha, (n_x, n_y)\rangle$ the one-particle state corresponding to a neutrino of flavor $\alpha$, with momentum $\vec{p} = (n_x, n_y)^T p_0$. We consider an initial quantum state where the three flavors appear, shown in blue in Fig.~\ref{fig:pgrid}. A simpler case, with only two initial neutrinos, is discussed in Appendix~\ref{app:2beams}. For our example, the electron neutrinos travel in the direction $- \vec{u}_y$, while the heavy-lepton flavor neutrinos travel in the direction $+ \vec{u}_y$, in the positive (negative) $\vec{u}_x$ direction for $\nu_\mu$ ($\nu_\tau$).
\begin{equation}
    \label{eq:initial_3beams}
|\psi_0\rangle = \hat{\mathcal{A}} \ |\nu_\tau,(-1,+1) \rangle \otimes |\nu_e,(0,-1)\rangle \otimes |\nu_\mu,(+1,+1) \rangle \, ,
\end{equation}
where $\hat{\mathcal{A}}$ is the antisymmetrization operator. In terms of one-body density matrices, this corresponds to the nonzero initial entries $\left(f_{(-1,+1)}\right)^\tau_\tau = 1$, $\left(f_{(0,-1)}\right)^e_e = 1$ and $\left(f_{(+1,+1)}\right)^\mu_\mu = 1$.

For mean-field and QKE calculations, we solve the differential equations~\eqref{eq:meanfield}--\eqref{eq:QKE} with the Scipy routine \texttt{solve\_ivp}~\cite{2020SciPy-NMeth}. For the many-body calculations, we perform state-vector simulations with Qiskit-Aer~\cite{Javadi-Abhari:2024kbf}.

\begin{figure*}[ht]
    \centering
    \includegraphics[width=0.9\textwidth]{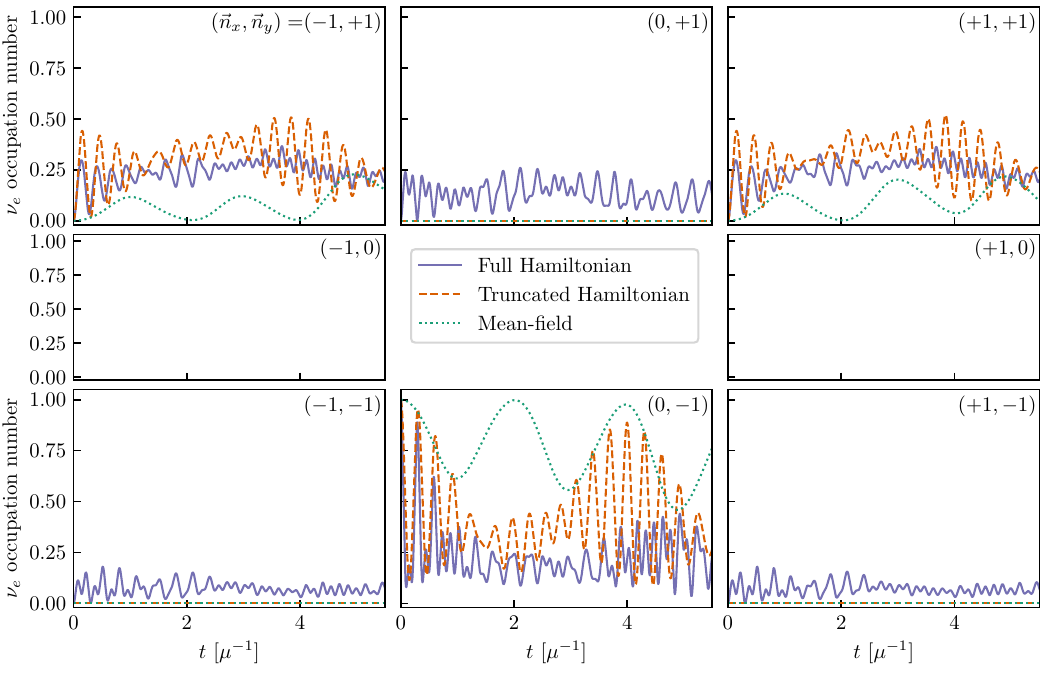}
    \caption{Time evolution of the $\nu_e$ occupation numbers for the initial state \eqref{eq:initial_3beams}. Many-body calculations using the full (forward/exchange) Hamiltonian are shown with solid (dashed) lines, while the mean-field result is shown with a dotted line. Each panel corresponds to the momentum point $(p_x,p_y)$ on the grid shown in Fig.~\ref{fig:pgrid}.}
    \label{fig:3beams}
\end{figure*}

\subsection{Many-body and mean-field evolutions}

\begin{figure*}[!ht]
    \centering
    \includegraphics[width=0.9\textwidth]{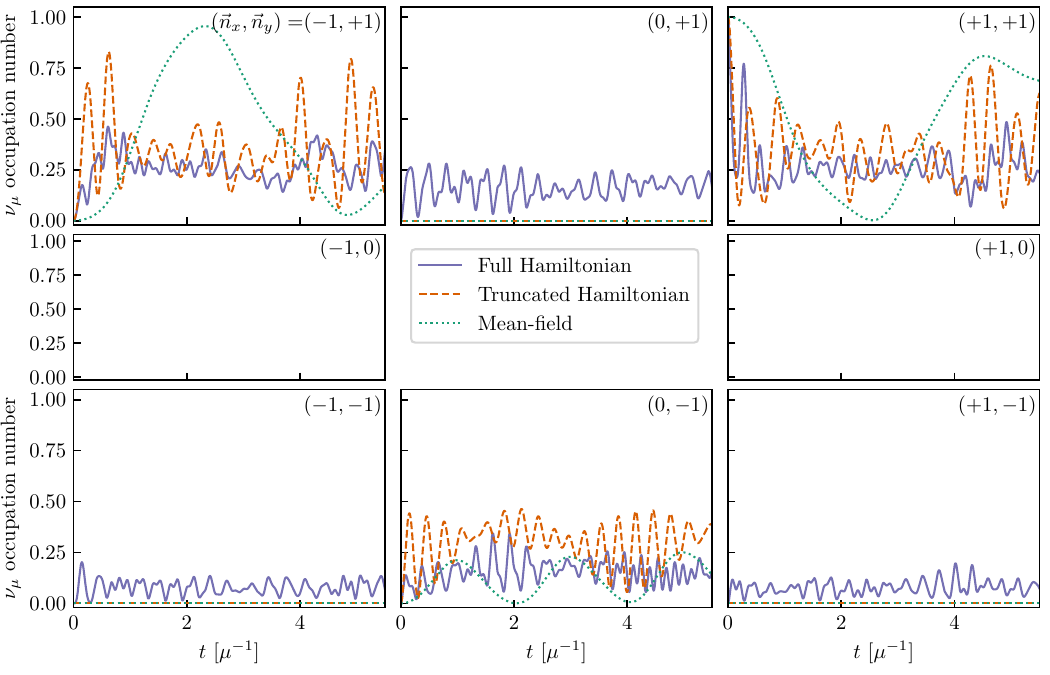}
    \caption{Same as Fig.~\ref{fig:3beams}, but for the $\nu_\mu$ occupation numbers.}
    \label{fig:3beams_numu}
    \vspace{1.15cm}
    \includegraphics[width=0.9\textwidth]{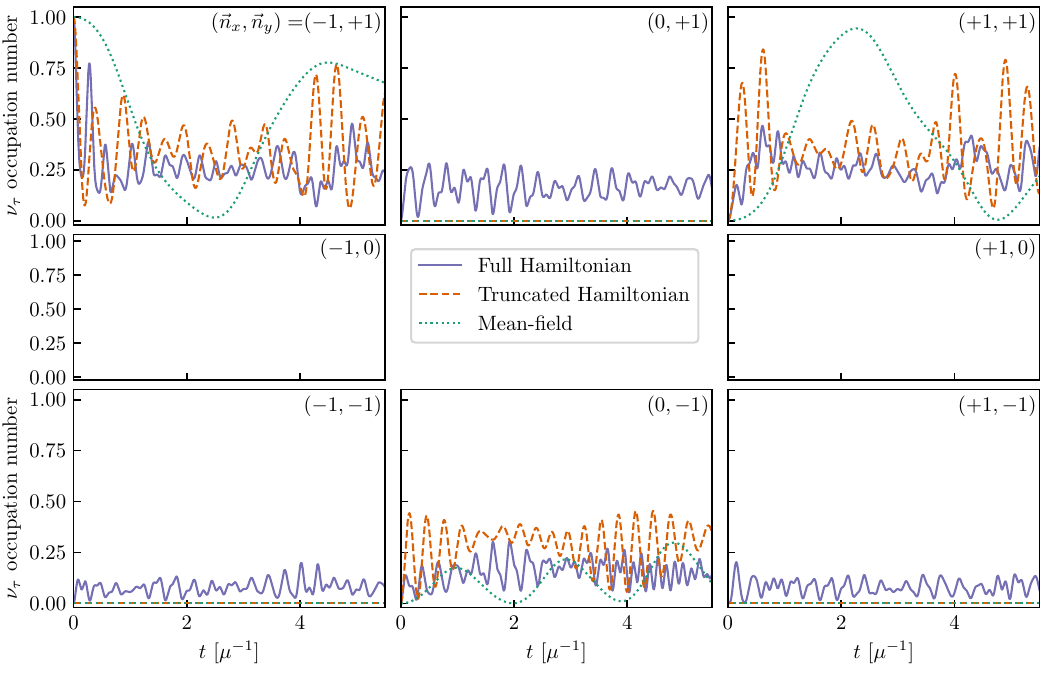}
    \caption{Same as Fig.~\ref{fig:3beams}, but for the $\nu_\tau$ occupation numbers.}
    \label{fig:3beams_nutau}
\end{figure*}

The differences in the evolution of the system between the many-body and mean-field calculations can readily be seen by looking at the occupation numbers of each one-particle state. We show in Fig.~\ref{fig:3beams} the evolution of these quantities for the electron flavor (see Figs.~\ref{fig:3beams_numu} and \ref{fig:3beams_nutau} for the muon and tau flavors). The different panels reproduce the momentum grid of Fig.~\ref{fig:pgrid}, and we compare the evolution of occupation numbers for the mean-field equation [Eq.~\eqref{eq:meanfield}, dotted green line], the ``full'' many-body evolution [Eq.~\eqref{eq:schrodi_full}, solid purple line], and the many-body evolution under the ``truncated'' Hamiltonian $H = H_\mathrm{vac} + H_{\nu \nu}^{\mathrm{(f/e)}}$ [Eq.~\eqref{eq:schrodi_fwd}, dashed orange line]. The time is measured in units of the self-interaction strength~\eqref{eq:self_strength}.

We observe very large differences between the mean-field and many-body results, with a much faster evolution for the latter. These differences are associated to a rapid growth of entanglement in the many-body cases (see Sec.~\ref{sec:quantumness}), and agree with the longstanding view in the literature that the mean-field approximation could miss some important features of neutrino evolution. Furthermore, we find in agreement with~\cite{Cirigliano:2024pnm} that there is no a priori justification for the restriction of the Hamiltonian~\eqref{eq:H_forward} widely used in the literature, apart for computational reasons. Indeed, the full Hamiltonian ``opens'' interaction channels which can occupy initially unoccupied momentum states, for instance, here the states $(\vec{n}_x, \vec{n}_y) = (-1,-1), \, (0,1) \text{ and } (1,-1)$. As noted in~\cite{Cirigliano:2024pnm}, the timescales of many-body flavor evolution and momentum redistribution are the same (here, $\sim 0.1 \, \mu^{-1}$), while the quantum kinetic picture suggests a large difference between the flavor oscillation and collisional timescales. We come back to this point in Sec.~\ref{subsec:3beam_QKE}.

\begin{figure}[!ht]
    \centering
    \includegraphics[width=0.93\linewidth]{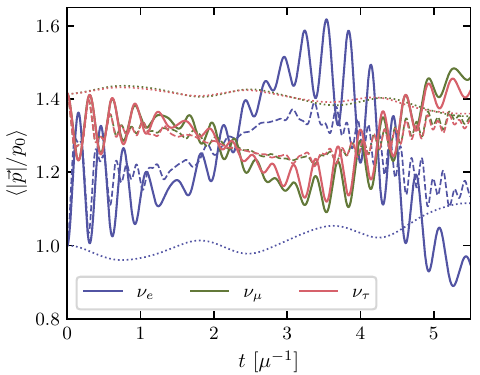}
    \caption{Evolution of the average kinetic energy for each neutrino flavor. Plotting conventions are the same as Fig.~\ref{fig:3beams}.}
    \label{fig:3beams_avE}
\end{figure}

The average momentum magnitude for each flavor is shown in Fig.~\ref{fig:3beams_avE}. As the neutrinos are ultrarelativistic, differences in momentum between mean-field and many-body correspond to differences in the energy of the detected neutrinos, with potential implications for a neutrino detection. Specifically, the mean-field calculations show a higher energy difference between electron and heavy lepton flavors at small times, while the many-body treatment brings these energies closer to each other. A lower energy asymmetry could imply a higher proton fraction in the ejecta~\cite{Fischer:2023ebq,Wang:2023tso}, although our model is far too simplistic to draw such conclusions with high confidence. A proper supernova simulation including these many-body aspects would address such possibilities, but it is outside the scope of this paper—see~\cite{Qiu:2025kgy,Qiu:2025ybw,Balantekin:2023ayx} for related studies in NSM and CCSN environments.

\subsection{Comparison with quantum kinetics}
\label{subsec:3beam_QKE}

Our results show significant differences between the truncated and full many-body approaches, themselves very different from the mean-field approach. Interestingly, the possibility of momentum transfer beyond the forward/exchange cases appears in the QKE formalism, but only at the collisional level. In order to compare the momentum distribution resulting from scatterings compared to the many-body evolution, we solve here the discrete QKE~\eqref{eq:QKE} for the same initial configuration~\eqref{eq:initial_3beams}.
There are two key differences between the full many-body and QKE evolutions, which we illustrate in the following. Firstly, there is a separation of timescales at the heart of the QKE formalism (with, broadly speaking, a $\mathcal{O}(G_F)$ term responsible for flavor transformation and a $\mathcal{O}(G_F^2)$ term responsible for momentum exchange), while the many-body timescales are conflated. Secondly, the QKE collision term populates some momentum bins which cannot be populated in the many-body calculation.

\subsubsection{Timescales}

\begin{figure}[!ht]
    \centering
    \includegraphics[width=0.95\linewidth]{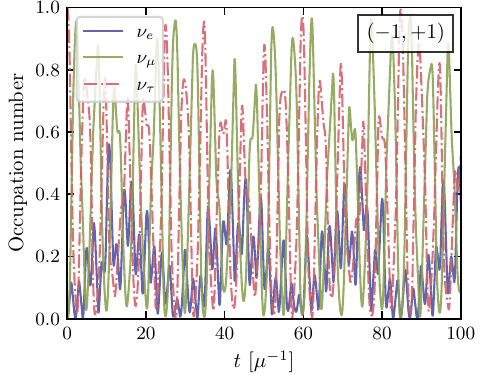}
    \caption{Large $t$ evolution of the occupation numbers for the momentum bin $(-p_0,p_0)$ in the mean-field approximation. This plot is an extension of the dotted lines of the top left panels of Figs.~\ref{fig:3beams}--\ref{fig:3beams_nutau}, showing a long-term quasi-periodicity.}
    \label{fig:3beams_MF_long}
\end{figure}

In the QKE case, the ratio between the scales of the collision and mean-field terms is [see Eqs.~\eqref{eq:discrete_collision} and \eqref{eq:QKE}]
\begin{equation}
\label{eq:ratio_coll_MF}
    \frac{\left(\frac{G_F}{V}\right)^2 V^{1/3}}{\frac{G_F}{V}} = \frac{G_F}{V^{2/3}} \sim 2.7 \times 10^{-14} \, .
\end{equation}
Although this ratio slightly underestimates the strength of the collision term (because the associated phase space, which we do not include, is larger in the numerator than the denominator), we are still confronted with a dramatic difference of timescales. We show in Fig.~\ref{fig:3beams_MF_long} the long-term evolution of one momentum mode of the system in the mean-field approximation, for $t$ up to $100 \mu^{-1}$. We see that the system settles in quasi-periodic oscillations, such that the collision term would slowly act on top of this relatively fast oscillatory behavior. Capturing this effect, but in an accelerated and numerically feasible way, can be achieved\footnote{One could also average over the short timescale in order to describe the slow evolution due to collisions, as done in other setups in, e.g.,~\cite{Froustey:2020mcq,Fiorillo:2023ajs}.} by rescaling the collision term with a factor $\kappa \gg 1$. Specifically, we take $\kappa = 3 \times 10^{10}$ for illustrative purposes. This maintains the scaling ``collision'' $\ll$ ``mean-field,'' but makes the problem numerically tractable. The evolution of the occupation numbers is shown in Fig.~\ref{fig:3beams_QKE}.

\subsubsection{Momentum distribution}

\begin{figure*}[!ht]
    \centering
    \includegraphics[width=0.9\textwidth]{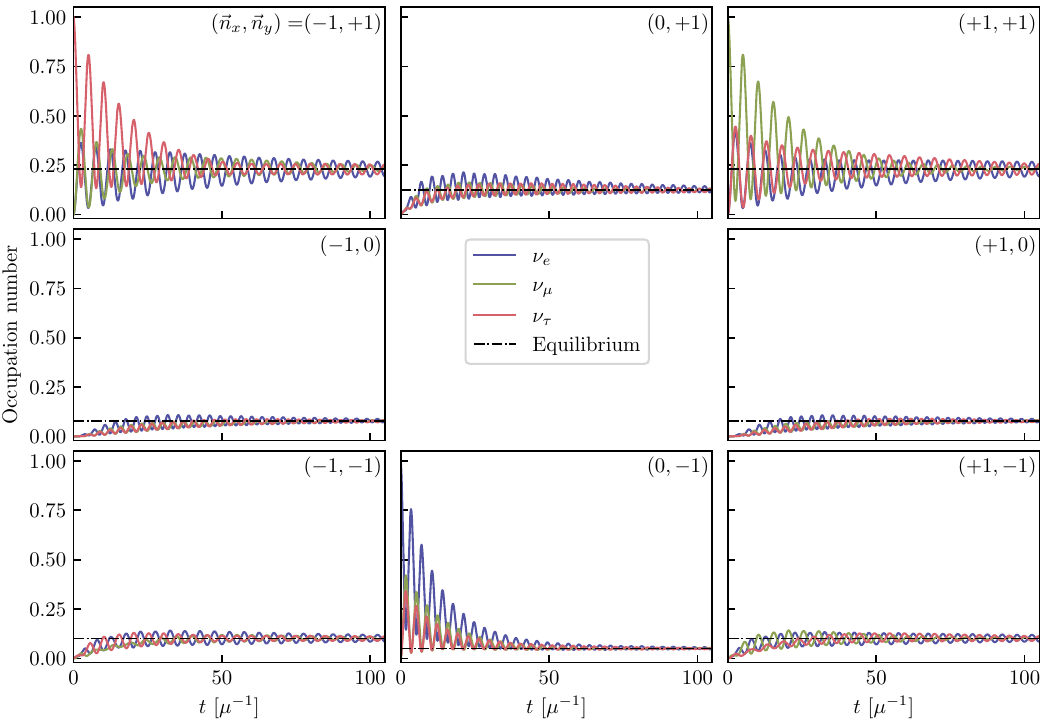}
    \caption{Time evolution of the occupation numbers for a discrete QKE calculation, to be compared with the many-body evolutions shown in Figs.~\ref{fig:3beams}--\ref{fig:3beams_nutau}. We emphasize that the collision term has been artificially multiplied by $\kappa = 3 \times 10^{10}$ to make its effect noticeable on a timescale comparable to the other calculations. The dash-dotted black line is the equilibrium prediction~\eqref{eq:eqb_QKE}, which in this particular case is the same for all three flavors. In reality, the same equilibrium would be reached after a much longer time $\sim 10 \, \kappa \, \mu^{-1}$.}
    \label{fig:3beams_QKE}
\end{figure*}

The set of momentum bins that are occupied by momentum-exchanging processes is different between the quantum kinetic and closed many-body calculations. This is visible in our example by looking at the panels representing the momenta $(-p_0,0)$ and $(+p_0,0)$ in Fig.~\ref{fig:3beams_QKE}, to be compared with the same panels in Figs.~\ref{fig:3beams}--\ref{fig:3beams_nutau}. Given our momentum grid, there is only one process satisfying momentum and energy conservation that can populate those bins from the others. Leaving aside the flavor degree of freedom, it corresponds to the momenta (in units of $p_0$)
\begin{equation}
\label{eq:indiv_process_1}
    \begin{pmatrix} 0 \\ -1 \end{pmatrix} + \begin{pmatrix} 0 \\ +1 \end{pmatrix} \longrightarrow \begin{pmatrix} -1 \\ 0 \end{pmatrix} + \begin{pmatrix} +1 \\ 0 \end{pmatrix} \, .
\end{equation}
Initially, the momentum bin $(0,-1)$ is occupied---by a pure $\nu_e$ state---but not $(0,+1)$. This latter bin gets populated through
\begin{equation}
\label{eq:indiv_process_2}
    \begin{pmatrix} 0 \\ -1 \end{pmatrix} + \begin{pmatrix} \pm 1 \\ +1 \end{pmatrix} \longrightarrow \begin{pmatrix} 0 \\ +1 \end{pmatrix} + \begin{pmatrix} \pm 1 \\ -1 \end{pmatrix} \, .
\end{equation}
When we directly evolve the $N$-body quantum state \eqref{eq:initial_3beams}, the process \eqref{eq:indiv_process_2} implies that the one-particle states $|\nu_\alpha, (0,+1)\rangle$ and  $|\nu_\beta, (0,-1)\rangle$ never coexist in the same $N$-body state. As a consequence, the process \eqref{eq:indiv_process_1} cannot be realized. However, in the QKE approach, we do not evolve a $N$-body wavevector, but a collection of occupation numbers. There is no entanglement and the momentum bins $(0,\pm 1)$ have simultaneously nonzero occupation number values. As a consequence, the process~\eqref{eq:indiv_process_1} can be realized---still, we recall that this occurs on a collisional timescale. Incidentally, since the process~\eqref{eq:indiv_process_1} first requires the occupation of the $(0,+1)$ state via~\eqref{eq:indiv_process_2}, the timescale on which the bins $(\pm 1, 0)$ are populated is larger than for the other bins, as is clearly seen in Fig.~\ref{fig:3beams_QKE}.

The overall evolution, when including the collision term, is a damping of mean-field oscillations towards an equilibrium state. We derive in Appendix~\ref{app:QKE} the prediction for this equilibrium momentum distribution, see Eq.~\eqref{eq:eqb_QKE}. It is obtained by maximizing the entropy functional in a generalized grand-canonical ensemble, under the constraints of global conservation of energy, momentum and particle number. It is depicted as a black dash-dotted line in Fig.~\ref{fig:3beams_QKE}, showing perfect agreement with the numerical solution.

\subsubsection{Discussion}

Since a many-body calculation using the full Hamiltonian allows for the population of new momenta states, while this is impossible in a mean-field calculation, the comparison of both frameworks will necessarily show significant differences. By comparing with a QKE calculation for the same setup, we explicitly showed that momentum redistribution is very different, mostly in terms of timescales.

In the QKE picture, flavor evolution due to the mean-field term can be very fast, while kinetic evolution is comparatively slow. Our results, in agreement with those of Ref.~\cite{Cirigliano:2024pnm}, suggest that momentum redistribution could occur much faster, on the same timescale as for flavor, in dense enough environments where many-body effects are relevant. We note, however, that we only focused on a very simple geometry with a small number of neutrinos.

The key question is therefore the domains of applicability of the many-body and QKE approaches. In particular, the derivation of the QKE collision term hinges upon a molecular chaos approximation, where two-body correlations are explicitly discarded between each scattering. This leads to the separation of timescales that is absent in our many-body calculation. This is expected, since our setup corresponds, \emph{in fine}, to ``plane waves in a box,'' which interact continuously and thus represent a very different regime from the one at the heart of the QKE picture (as pointed out in~\cite{Shalgar:2023ooi, Johns:2023ewj}).  For a regime where our many-body calculations describe more faithfully the system, it is then clear that using a truncated Hamiltonian is not justified at all. It is thus all the more crucial to develop more complex many-body calculations with the full Hamiltonian, which is a natural quantum computing problem. In the following section, we discuss the differences in ``quantumness'' between the truncated and full implementations, and estimate the cost of quantum computations of the time evolution as a function of the number of neutrinos involved. 

\section{Non-classicality of neutrino evolution}
\label{sec:quantumness}

We have shown in Sec.~\ref{sec:time_evolution} that the individual occupation numbers vary significantly depending on the evolution equation adopted. To further understand the difference between mean-field results and the many-body treatments, in this section we study entanglement and Ren{\'y}i entropies as measures of deviation from classical results (see Sec.~\ref{subsec:entanglement}).

If the simple many-body calculations we have carried out faithfully represent the conditions in some regions of dense astrophysical environments, our results clearly show, in agreement with Ref.~\cite{Cirigliano:2024pnm}, that using a truncated Hamiltonian is not justified. 
Unfortunately, $N$-body calculations of the Schrödinger equation are very computationally expensive problems, which naturally leads to the use of quantum computing hardware. Given the current state of the hardware, here we estimate the quantum resources needed for long term simulations. To that end, we first compute an upper bound to the Trotter error, then proceed to estimate the number of entangling gates $CZ$ and non-Clifford gates $R_Z$, required for a single Trotter step (see Sec.~\ref{subsec:Trotter}). Both these resources are computationally expensive, $CZ$ will give rise to many-body correlations, and $R_Z$ is the source of ``magic'' in the time evolution, as without non-Clifford gates simulations could be computed efficiently by classical computers~\cite{Aaronson:2004xuh}.

\begin{figure*}[!ht]
    \centering
\includegraphics[width=0.95\textwidth]{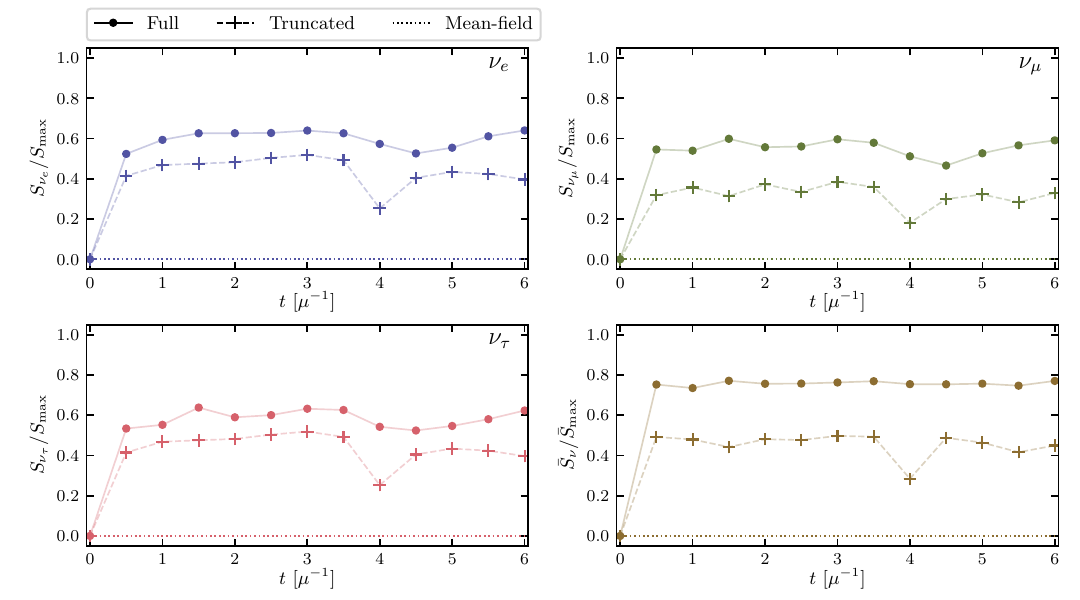}
    \caption{Entanglement entropy at selected time intervals, starting from the initial state~\eqref{eq:initial_3beams}. It is given by~\eqref{eq:entropy_flavor} for each neutrino flavor ($\nu_e$, top left; $\nu_\mu$, top right; $\nu_\tau$, bottom left), and by~\eqref{eq:entropy_onemode} for the average over one-particle modes (bottom right panel). Dots (resp.~crosses) denote the full (resp.~truncated) many-body results. In the mean-field calculation, the entanglement entropy is always zero.
    }
    \label{fig:3beams_entanglement}
\end{figure*}

\subsection{Entanglement and Stabilizer R\'{e}nyi Entropies}
\label{subsec:entanglement}

To quantify the deviation of the system from a mean-field configuration, we need metrics that are exactly zero for the mean-field wavefunction and increase with the difference with the many-body treatment. Entanglement entropy between neutrino modes is one such metric, which quantifies how entangled are the quantum numbers of the individual neutrinos. As the particle number increases with system size (number of momenta modes), we need to decide how to partition the system. For simplicity of analysis, we categorize all momenta modes into the different flavors, introducing for instance the reduced density matrix for the electron flavor $\rho_{\nu_e}(t)=\mathrm{Tr}_{\nu_\mu, \nu_\tau} \rho(t)$, where $\rho(t)= | \Psi(t) \rangle \langle \Psi(t) |$. Then, the entanglement entropy for the electron flavor can be written as,
\begin{equation}
\label{eq:entropy_flavor}
    S_{\nu_e}(t) \equiv - \mathrm{Tr} \left( \rho_{\nu_e} \ln \rho_{\nu_e} \right) \, ,
\end{equation}
and likewise for the other flavors. 

We also consider the average entanglement entropy of a neutrino mode with definite flavor and momentum, 
\begin{equation}
\label{eq:entropy_onemode}
\bar{S}_{\nu}= -N^{-1}_{\nu} \sum_{\alpha,\vec{p}} [\mathrm{Tr}  (\rho_{\nu_{\alpha,\vec{p}}} \ln \rho_{\nu_{\alpha,\vec{p}}})]    \, ,
\end{equation}
where the single mode density matrix $\rho_{\nu_{\alpha,\vec{p}}}$ is the result of tracing over all other modes in the system.

In Fig.~\ref{fig:3beams_entanglement} we depict $S_{\nu_\alpha}$ for all three flavors, and $\bar{S}_{\nu}$, as function of time for both the full and truncated Hamiltonian, re-scaled by the maximal values they can reach ($S_\mathrm{max}$, $\bar{S}_\mathrm{max}$). 
We notice that entanglement for the full Hamiltonian is consistently higher, and as a result, the wavefunction is qualitatively further distant from the mean-field predictions. When analyzing the entanglement effects based only on the flavor degree of freedom, the maximal value of the entropy is 
$S_\mathrm{max} = \ln(2^8)$, since there are 8 momenta modes per flavor, while maximal entropy for a single flavor and momentum mode is  $\bar{S}_\mathrm{max} = \ln(2)$.  In the mean-field limit, the reduced density matrices entering Eqs.~\eqref{eq:entropy_flavor}--\eqref{eq:entropy_onemode} correspond to pure state density matrices, for which the entropy vanishes.

\begin{figure}[!ht]
    \centering
    \includegraphics[width=0.94\columnwidth]{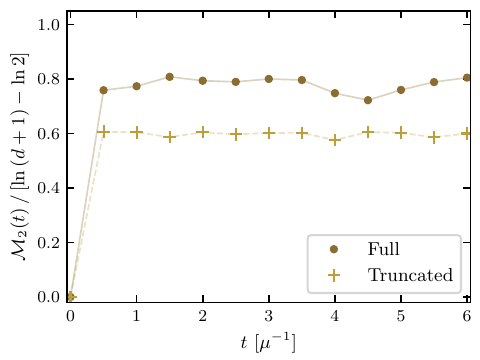}
    \caption{Stabilizer-2 Rényi entropy~\eqref{eq:M2} at selected time intervals for the initial configuration~\eqref{eq:initial_3beams}. In both cases we use 24 qubits to perform simulations ($d=2^{24}$), which corresponds to 3 flavors on a $3 \times 3$ two-dimensional momentum lattice (the origin being excluded to describe ultrarelativistic neutrinos).
    }
    \label{fig:3beams_magic}
\end{figure}

Another useful metric is the second moment R\'{e}nyi entropy ($\mathcal{M}_2$)~\cite{PhysRevLett.128.050402}, which is associated with the probability of the state being represented by a given Pauli
string. In simple terms, this metric quantifies how hard it would be to simulate the state using only stabilizer states which are efficiently representable by classical computers. The first quantitative study of this quantity in dense neutrino systems, although described by the truncated Hamiltonian, was performed in~\cite{Chernyshev:2024pqy}. Another colloquial term is ``magic'' resources, e.g. T gates,  and $\mathcal{M}_2$ is related to a lower bound on the number of such gates needed for a computation~\cite{Hou:2026}. The respective expression is,
\begin{equation}
\label{eq:M2}
\mathcal{M}_2(t) = -\ln\ d^{-1} \sum_{P  \in \mathcal{P}} \langle \Psi(t)|P| \Psi(t) \rangle^4
\end{equation}
where $d$ is the dimension of the Hilbert space and $\mathcal{P}$ is the set of all the possible Pauli strings for the given system size (e.g., $XZYIYYX$...). 
An upper bound on the stabilizer 2-Rényi entropy is given by $\mathcal{M}_2 < \ln(d+1) - \ln 2$~\cite{PhysRevLett.128.050402}. 
In Fig.~\ref{fig:3beams_magic}, we plot the evolution of the ratio of $\mathcal{M}_2$ with this upper bound, for the same times as the entanglement entropy in Fig.~\ref{fig:3beams_entanglement}. This bound may not be as tight as one would wish for specific cases~\cite{Chernyshev:2024pqy,Liu:2025frx}.
As the plot shows, the full Hamiltonian leads to an increase by 20\% of the Rényi entropy with respect to the upper bound. This makes the computations with the full Hamiltonian harder for classical methods, further increasing the divide between the two methods.

\subsection{Trotter error estimate and gate resources}
\label{subsec:Trotter}

For simplicity, we consider the first order Trotter step for the time evolution in the mass basis~\cite{Spagnoli:2025etu,Amitrano:2022yyn}. Our goal is to obtain upper bounds and, more precisely, the scaling of these bounds with the number of one-particle modes included in the calculation.

Breaking the time evolution operator into its one-body and two-body pieces leads to the error~\cite{Childs:2019hts}
\begin{equation}
\begin{aligned}
    \left \lVert e^{- \i H t} - e^{- \i H_\mathrm{vac} t} e^{- \i H_{\nu \nu}t} \right \rVert &\leq \frac{t^2}{2} \left \lVert \left[H_\mathrm{vac}, H_{\nu \nu}\right] \right \rVert \\
    &\equiv \frac{t^2}{2} \left \lVert C_{12} \right \rVert \, .
\end{aligned}
\end{equation}
Further sources of error come from breaking the vacuum and interaction terms into individual components. The one-body term $H_\mathrm{vac}$, written in the mass basis, is a sum of commuting one-body operators. As such it can be exactly further decomposed as
\begin{equation}
    e^{-i H_\mathrm{vac} t} = \prod_{\alpha, \beta} \prod_{i=1}^{N}{e^{-\i \omega_i \had_{\alpha,i}\ha_{\alpha,i}}} \, ,
\end{equation}
without incurring additional errors.

The two-body term, defined in Eq.~\eqref{eq:H_full_discrete}, can be written as $H_{\nu \nu} = \mu \sum_{K=1}^{\Gamma} h_K$. The decomposition of the time evolution operator for $H_{\nu \nu}$ is associated with the first order Trotter error~\cite{Childs:2019hts,Amitrano:2022yyn}
\begin{equation}
    \left \lVert e^{- \i H_{\nu \nu} t} - \prod_{K=1}^{\Gamma}{e^{- \i \mu h_K}} \right \rVert \leq \frac{t^2}{2} \mu^2 \sum_{K=1}^{\Gamma} \left\lVert \sum_{L = K+1}^{\Gamma} \left[h_K, h_L\right] \right\rVert \, ,
\end{equation}
and we further bound the right-hand side using the triangular inequality to get
\begin{align}
    \left \lVert e^{- \i H_{\nu \nu} t} - \prod_{K=1}^{\Gamma}{e^{- \i \mu h_K}} \right \rVert &\leq \frac{t^2}{2} \mu^2 \sum_{K < L} \left\lVert  \left[h_K, h_L\right] \right\rVert \nonumber \\
    &= \frac{t^2}{2} \frac{\mu^2}{2} \sum_{K \neq L} \left\lVert  \left[h_K, h_L\right] \right\rVert \nonumber \\
    &\equiv \frac{t^2}{2} \lVert C_{22} \rVert \, . \label{eq:def_C22}
\end{align}

\subsubsection{Splitting one-body/two-body}

First, we compute an upper bound on $C_{12}$ by separating the forward/exchange part and the other contributions.

\paragraph*{Forward/exchange part} We want to compute
\begin{equation}
    \label{eq:C12_forward_def}
    C_{12}^\mathrm{(f/e)} = \left[H_\mathrm{vac},H_{\nu \nu}^{\mathrm{(f/e)}}\right] \, .
\end{equation}
Details are given in Appendix~\ref{subsec:C12}, and we obtain the upper bound:
\begin{equation}
\label{eq:C12_forward}
    \left \lVert C_{12}^\mathrm{(f/e)}\right \rVert \leq 8 \mu \, \max_{\alpha, i, j}{\lvert \omega_{\alpha i} - \omega_{\alpha j}\rvert} \times N_F(N_F-1) \times N(N-1) \, .
\end{equation}
We emphasize that $N$ is not the number of neutrinos, but the number of momentum modes in the grid.\footnote{This would be the same for a calculation with the truncated Hamiltonian, but here we focus on the forward/exchange contribution to the Trotter error for a full Hamiltonian calculation.} This result is similar to the estimate in \cite{Spagnoli:2025etu} [see their Eq.~(B7)], noting that $\mu_{\text{\cite{Spagnoli:2025etu}}} \propto N \mu$, and the quantities $\omega$ have slightly different definitions. Also, we note that the result~\eqref{eq:C12_forward} is compatible with the results in \cite{Amitrano:2022yyn} (in which $C_{12} = 0$), because their calculation is for a single-energy neutrino gas (with different angles), and $\omega_{\alpha i}$ only depends on the magnitude of momenta.

\paragraph*{Non-forward part} We can compute similarly the contribution
\begin{equation}
    C_{12}^\stkout{\mathrm{(f/e)}} = \left[H_\mathrm{vac},H_{\nu \nu} - H_{\nu \nu}^{\mathrm{(f/e)}}\right] \, .
\end{equation}
We refer once more to Appendix~\ref{subsec:C12} for the details. We obtain the following bound:
\begin{multline}
    \label{eq:C12_nonforward}
    \left \lVert  C_{12}^\stkout{\mathrm{(f/e)}} \right \rVert \leq 32 \mu \, \max_{\alpha, i, j}{\lvert \omega_{\alpha i} - \omega_{\alpha j}\rvert} \times N_F^2 \\
    \times N(N-1) \times N_\mathrm{PS} \, ,
\end{multline}
where we bounded $\mathcal{V}$ by 16, which is obvious from the expression~\eqref{eq:V_spherical}. Note that, if we are in two dimensions, we have $| \mathcal{V} | \leq 4$ (as can be seen from \eqref{eq:V_spherical}, setting all $\theta_i = \pi/2$), such that the above bound would be four times tighter. We noted $N_\mathrm{PS}$ the maximum number of points $(i_3, i_4)$ on the grid which satisfy both kinetic energy and momentum conservation. For the small momenta grid we consider in this work, it is a number of order unity which almost does not vary with the grid size.

If one could use a highly resolved momentum grid, we could obtain a scaling of $N_\mathrm{PS}$ in the continuum limit. Calling $n_\mathrm{max}$ the index associated to the largest momentum in the grid, such that the “UV-cutoff” of our model is $(2 \pi/L) n_\mathrm{max} = \Lambda$, we have in the continuum limit $N=\frac43 \pi n_\mathrm{max}^3$ momentum modes. Following the calculations of Appendix~\ref{app:phase_space}, and specifically Eq.~\eqref{eq:scaling_NPS}, we find that for very large $N$ one would have the scaling $N_\mathrm{PS} \propto N^{2/3}$, with a multiplicative factor of order unity.

\subsubsection{Two-body operator splitting}

We have introduced the first order Trotter error associated with the two-body operator in~\eqref{eq:def_C22}, and it reads explicitly

\begin{equation}
\label{eq:C22_def}
\begin{split}
   \lVert C_{22} \rVert =& \frac12 \mu^2 \sum_{\alpha, \beta, \gamma,\sigma} \sum_{i_1 \neq i_2, j_1 \neq j_2} \sum_{i_3, j_3} |\mathcal{V}_{i_1 \dots} \mathcal{V}_{j_1 \dots}| \\
   &\times \left\lVert\left[\had_{\alpha,i_1} \ha_{\alpha,i_3} \had_{\beta,i_2} \ha_{\beta,i_4}, \had_{\gamma,j_1} \ha_{\gamma,j_3} \had_{\sigma,j_2} \ha_{\sigma,j_4}\right] \right\rVert \, , 
\end{split}
\end{equation}
using the shorthand notation $\mathcal{V}_{i_1 \dots} = \mathcal{V}(\vec{p}_{i_1},\vec{p}_{i_2},\vec{p}_{i_3},\vec{p}_{i_4})$. The details are given in Appendix~\ref{app:commutator}, and we obtain
\begin{equation}
\lVert C_{22} \rVert \leq 2^{10} \mu^2 \times N_F^3 \times N^3 \times N_\mathrm{PS}^2 \, ,
\end{equation}
which would scale in the continuum limit as $N_F^3 N^{13/3}$.

In the forward/exchange only case, the bound on $\lVert C_{22}^{(\mathrm{f/e})}\rVert$ scales as $N^3$, consistent with~\cite{Spagnoli:2025etu}.

\begin{figure*}[!ht]
    \centering    \includegraphics[width=0.85\textwidth]{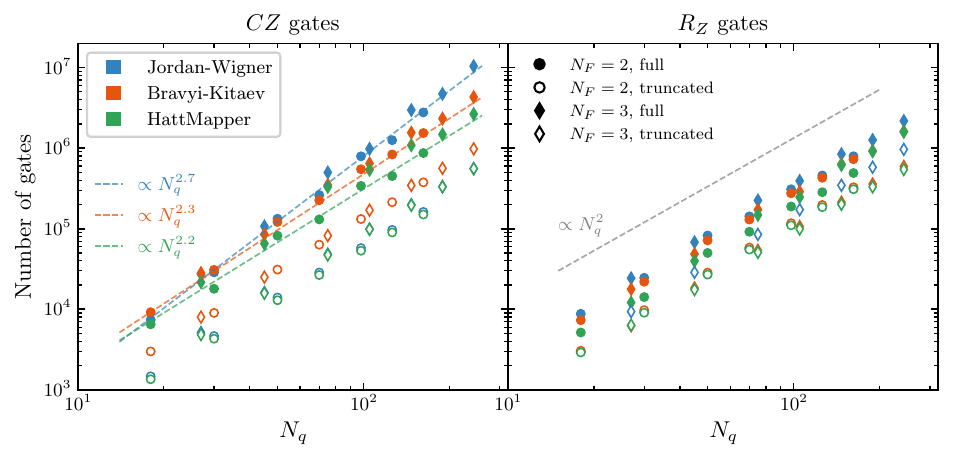}
    \caption{Number of $CZ$ (left) and $R_Z$ (right) gates for a single Trotter step, for 2 (circle) and 3 (diamond) flavors for the full and truncated Hamiltonian as a function of the system size (number of qubits, $N_q$). We display the results obtained with the Jordan-Wigner (blue), Bravyi-Kitaev (orange), and HattMapper (green) fermion-to-qubit transformations. The number of $R_Z$ gates scales quadratically with the number of qubits (gray line on the right panel), while numerical fits provide the scaling of the number of $CZ$ gates (for the full Hamiltonian) on the left panel, with different results depending on the encoding.}
    \label{fig:quantum_resources}
\end{figure*}

\subsubsection{Quantum Resources}

Provided the Trotter error has been computed, we turn our attention to the quantum computational cost of a single first order Trotter step. At first we need to represent the second quantized Hamiltonian in the qubit basis. To this end, we employ the standard Jordan-Wigner (JW)~\cite{Jordan:1928wi} and Bravyi-Kitaev (BK)~\cite{Bravyi:2000vfj} transformations, and more recent work we conducted on numerically optimized mapping, called HattMapper (HM)~\cite{Liu:2024rtx}. It utilizes ternary tree mapping and a bottom-up construction procedure to generate Hamiltonian aware Fermion-to-qubit mapping to reduce the Pauli weight of the qubit Hamiltonian (see also~\cite{Li:2022oly,Liu:2024vbw} for other mapping methods). For illustration purposes, we focus on the two dimensional geometry with lattice sizes $n_x,n_y\leq 4$ for two and three neutrino flavors, both truncated and full Hamiltonian in Fig.~\ref{fig:quantum_resources}. The transpilation of the respective circuits into $(CZ,R_Z,H,S,S^\dagger)$ basis was performed with Qiskit~\cite{Javadi-Abhari:2024kbf}. The number of qubits is $N_q= N_F \times N$, where $N_F$ is the number of flavors and $N = (2n_x-1) (2n_y-1)$ is the number of momenta modes.

As expected, the resources for the full Hamiltonian are significantly higher than for the truncated counterpart, in agreement with the number of terms in the Hamiltonians. As we are implementing a first order Trotter step, the asymptotic scaling of the number of $R_Z$ gates is proportional to the number of the Pauli terms in the Hamiltonian, namely, $\sim (N_{PS}) N_q^2$, with the prefactor $N_\mathrm{PS}$ for the full Hamiltonian.\footnote{As we mentioned after Eq.~\eqref{eq:C12_nonforward}, for the small momenta grid considered here, $N_\mathrm{PS}$ is of order unity and does not vary significantly between our calculations. We thus expect a $\propto N_q^2$ scaling of the number of $R_Z$ gates in both the truncated and full cases.} These gates contain all the non-Clifford operations (``magic'') and are expected to have the highest computational cost for fault tolerant quantum computers. Note that in Fig.~\ref{fig:3beams_magic} we computed $\mathcal{M}_2$ as a proxy for these resources, while we here compute the actual resources needed for specific system sizes and geometries.
They are decomposed into a high number of T-gates~\cite{Selinger:2012pqc} which require magic factories at the hardware level~\cite{Ding:2018}. Since $\mathcal{M}_2$ is a log quantity, the difference between the full and truncated Hamiltonians is more visible in the gate count, and the trends agree. The choice of mapping has non-trivial implications for the numerical prefactor in the asymptotic scaling, and can vary the actual resources by up to an order of magnitude for the system Hamiltonian and grid of momenta modes. 

The number of $CZ$ gates, responsible for entanglement, is expected to be higher than the $R_Z$ gates, as Pauli gadgets comprised of $CZ$ sequences are used to diagonalize the various Pauli terms in the Hamiltonian. However, this is a rather inefficient method to decompose  Pauli exponentials into native quantum gates, and transpilation is an active area of research. From Fig.~\ref{fig:quantum_resources} we can see the impact of the Fermion-to-qubit mappings on the number of $CZ$ gates and the asymptotic scaling as function of system size. 
In the case of the forward Hamiltonian, BK results in a larger number of $CZ$ gates with respect to JW, but lower number of rotations $R_Z$. HM matches the best in either case, providing optimal results for both type of resources. In the case of the full Hamiltonian, JW under performs. As system size increases, the discrepancy between BK and HM, in the number of $R_Z$ gates, decreases, with HM again providing the optimal number of resources. Overall, HM provides the most resource efficient encoding for both types of resources. Assuming a large number of Trotter steps is required for the time evolution, e.g. $10^3$, the overall resources needed for systems of 100 modes are of the order of $10^{10}$. These resources are on the low end for typical high-energy problems and comparable to (slightly higher than) quantum chemistry problems—see figures in section IV in~\cite{osti_2588210}. 

\section{Summary and Conclusion}
\label{sec:conclusion}

The evolution of neutrinos in dense astrophysical environments is inherently a quantum many-body problem. While many studies have focused on the physics of flavor oscillations from a many-body perspective, they generally use a simplified (``truncated'') Hamiltonian, where only forward scattering two-body terms are considered. This is not a priori justified: in this work, we study simple neutrino systems on a momentum lattice, showing significant differences between many-body calculations using the full Hamiltonian, or a truncated version. In the wake of the study by Ref.~\cite{Cirigliano:2024pnm}, which first incorporated the full neutrino Hamiltonian into a many-body Schrödinger equation, we investigate the role of non-forward processes along several new directions.

First, we compare the evolution of one-body observables (occupation numbers) between the mean-field, truncated and full many-body frameworks (see Figs.~\ref{fig:3beams}--\ref{fig:3beams_nutau}). There are very large discrepancies between the mean-field and many-body results, and also clear differences between the two many-body calculations—the most visible being the population of new momentum states, as enabled by the non-forward Hamiltonian terms. Since this kinetic redistribution is impossible in the mean-field approach, we also implement a discrete version of the QKE collision term, in order to compare the momenta distributions between the QKE and many-body approaches (see Fig.~\ref{fig:3beams_QKE}). There are two major differences, discussed in Sec.~\ref{subsec:3beam_QKE}. First, in the quantum kinetic picture, momentum-changing processes correspond to rare individual scatterings, with an in-between ``loss of memory'' (molecular chaos), which leads to a separation of timescales between the flavor evolution (driven by the $\mathcal{O}(G_F)$ mean-field potential) and the kinetic evolution (driven by the $\mathcal{O}(G_F^2)$ collision term). In our many-body setup, there is no such separation and, as previously observed in~\cite{Cirigliano:2024pnm}, flavor and kinetic evolutions occur on similar timescales. The second difference is a consequence of the molecular chaos approximation: since no correlations are retained in the QKE picture, the system is entirely described by the one-body occupation numbers. This enables some scatterings which are impossible in our many-body calculation because they would need, in the initial state, two simultaneously occupied momenta which do not appear together in the $N$-body state. As a consequence, some momentum occupation numbers are nonzero in the QKE equilibrium (which we can predict analytically in a generalized grand-canonical ensemble, see Appendix~\ref{app:QKE}), but vanish in the many-body calculation.

In agreement with the criticism raised in Ref.~\cite{Johns:2023ewj} against an arbitrary truncation of the Hamiltonian, we find that under conditions where many-body effects become important, one cannot restrict the Hamiltonian to its forward/exchange part. We note that addressing that problem does not reduce the disagreement with mean-field results: the discrepancy increases when the full Hamiltonian is considered, precisely because of the non-forward interactions. 

We have quantified these non-mean-field features by studying various measures of the entanglement of the system and of the non-classicality (magic) of the evolution, see Sec.~\ref{subsec:entanglement}. All these quantities are increased with the full Hamiltonian, which signals a dramatic \emph{quantum} computational challenge. We thus determine in Sec.~\ref{subsec:Trotter} a bound on the first-order Trotter error associated with one time step, and its scaling with the size of the momentum grid $N$, with results in the truncated case consistent with~\cite{Spagnoli:2025etu}. The error incurred with the full Hamiltonian is increased by the presence of its additional terms, namely a phase-space contribution. While in the limited setups we have considered, this does not significantly change how the error scales with $N$, for very resolved grids (which would be the ultimate goal) it does so by several powers.

We finally analyze the quantum-resource cost of a single Trotter step, focusing on the number of $CZ$ and $R_Z$ gates obtained after transpilation of both the truncated and full Hamiltonians (see Fig.~\ref{fig:quantum_resources}). Our results highlight the very large gate counts required for this problem, the further increase induced by the inclusion of the non-forward terms, and the potential for substantial improvements through a more adequate fermion-to-qubit encoding. For instance, the HattMapper mapping developed in~\cite{Liu:2024rtx} outperforms the standard Jordan-Wigner and Bravyi-Kitaev transformations.

The setup we use (neutrinos in definite momentum states on a grid, aka ``plane waves in a box'') was also criticized in~\cite{Shalgar:2023ooi}, since it would describe a situation where neutrinos interact ``too much'' compared with actual environments, where the separation of scales at the heart of the QKE picture is valid. One can see the QKE and many-body calculations presented in this work as the two extremities of a spectrum: no correlations in the former, maximal correlations in the latter. A complete understanding of neutrino evolution and transport requires understanding the intermediate conditions, which in turn requires improving our description of the limiting regimes. Several approaches have been recently developed in the literature to bridge the gap between those two extremes, such as a semi-classical framework for non-forward processes~\cite{Carlson:2026mir}, wavepacket treatments~\cite{Cervia:2025pfg}, so-called once-in-a-lifetime encounter models~\cite{Kost:2024esc,Kost:2025vyt}, or inhomogeneous tensor network frameworks~\cite{Laraib:2025uza,Laraib:2025ziz}. However, a complete and satisfying treatment of collective neutrino flavor oscillations will require further developments in both the theoretical and computational avenues.

\begin{acknowledgments}

We thank Yukari Yamauchi for useful discussions on many-body calculations and for comparing our Hamiltonian expressions.

J.F. acknowledges support from the Severo Ochoa Excellence Grant CEX2023-001292-S funded by MICIU/AEI/10.13039/501100011033 and from the Network for Neutrinos, Nuclear Astrophysics and Symmetries (N3AS), through the National Science Foundation Physics Frontier Center Award No.~PHY-2020275.
YL and GL are supported in part by National Science Foundation CAREER Award No.~CCF-2338773 and ExpandQISE Award No.~OSI-2427020. GL is also supported by the Intel Rising Star Faculty Award. This work was supported by the U.S. Department of Energy, Office of Science, Office of Advanced Scientific Computing Research under Contract No. DE-AC05-00OR22725 through the Accelerated Research in Quantum Computing Program MACH-Q project. ER is supported by the U.S. Department of Energy (DOE) under Contract No. DE-AC02-05CH11231, through the National Energy Research Scientific Computing Center (NERSC), an Office
of Science User Facility located at Lawrence Berkeley National Laboratory. 
VC acknowledges support from the U.S. DOE under Grant No. DE-FG02-00ER41132. 

\end{acknowledgments}

\bibliography{biblio}

\clearpage

\onecolumngrid
\appendix

\appsection{Conventions and neutrino Hamiltonian}
\label{app:deriv_H}

In this Appendix, we specify our conventions for the spinors and the associated expression of the weak interaction matrix elements, leading to the expression for the Hamiltonian~\eqref{eq:H_full}.

The neutral-current Hamiltonian, restricted to the neutrino interaction part, reads (see, e.g.,~\cite{Volpe:2013uxl})
\begin{equation}
    H_{\nu \nu} =  \frac{G_F}{4 \sqrt{2}} \sum_{\alpha,\beta} \int{\dd^3{\vec{x}} \left[\bar{\psi}_{\nu_\alpha}\gamma_{\mu}(1-\gamma_5)\psi_{\nu_\alpha}\right]\left[\bar{\psi}_{\nu_\beta}\gamma^{\mu}(1-\gamma_5)\psi_{\nu_\beta}\right]} \, .
\end{equation}

\subsection{Spinor conventions}

We follow the conventions of~\cite{Peskin:1995ev}, with the expansion of the Dirac field in Fourier space
\begin{equation}
        \label{eq:psi_Peskin}
    \psi_{\nu_\alpha}(\vec{x}) = \sum_{h} \int{\frac{\dd^3{\vec{p}}}{(2 \pi)^3 \sqrt{2 E_p}} \left[\ha_{\alpha}(\vec{p},h) u_{\vec{p}}^{(h)} e^{i \vec{p} \cdot \vec{x}} + \hbd_{\alpha}(\vec{p},h) v_{\vec{p}}^{(h)} e^{- i \vec{p} \cdot \vec{x}}\right]} \, .
\end{equation}
The creation/annihilation operators satisfy the anticommutation rule  $\{\ha_{\alpha}(\vec{p},h), \had_{\beta}(\vec{q},h')\} = (2 \pi)^3 \,  \delta^{(3)}(\vec{p} - \vec{q}) \, \delta_{\alpha \beta} \delta_{h h'}$, and likewise for $\hb$, $\hbd$. In the following, we will restrict to the case of a gas of ultrarelativistic left-handed neutrinos, such that we only consider the $\ha$, $\had$ operators with negative helicity, with the shorthand notation $\ha(\vec{p}) = \ha(\vec{p},-)$ and the anticommutation relation~\eqref{eq:had_ha}. The energy is also $E_p = p$.

In the Weyl basis, the gamma matrices are
\begin{equation}
    \label{eq:gamma_matrices}
\gamma^0= \begin{pmatrix}
            0 && 1 \\
            1 && 0
           \end{pmatrix} \ , \quad
 \gamma^i= \begin{pmatrix}
            0 &&  \sigma^i \\
            -\sigma^i && 0
           \end{pmatrix} \ , \quad
\gamma^5=\begin{pmatrix}
            -1 && 0 \\
            0 && 1
           \end{pmatrix} \, ,
\end{equation}
where $\sigma^i$ are the Pauli matrices, which can be arranged in a vector $\vec{\sigma} = (\sigma^x, \sigma^y, \sigma^z)$. The massless left-handed spinor reads
\begin{equation}
\label{eq:spinors}
u^{(-)}_{\vec{p}}= \frac{p}{\sqrt{p-p_z}}
                    \begin{pmatrix}
                     (1-\vec{\sigma} \cdot \hat{\vec{p}})\cdot \begin{pmatrix}1 \\ 0 \end{pmatrix}\\ 
                     \vec{0}\\
                    \end{pmatrix}
                = \frac{p}{\sqrt{p-p_z}} \begin{pmatrix} 1- \hat{p}_z \\ -(\hat{p}_x + \i \hat{p}_y) \\ 0 \\ 0 \end{pmatrix} \, , \qquad \text{with } \hat{\vec{p}} \equiv \vec{p}/\lvert \vec{p}\rvert \, .
\end{equation}

With the expansion~\eqref{eq:psi_Peskin}, the self-interaction Hamiltonian is written
\begin{multline}
    H_{\nu \nu} = \sqrt{2} G_F \sum_{\alpha,\beta} \int \frac{\dd^3 \vec{p}_1}{(2 \pi)^3 \sqrt{2 p_1}} \frac{\dd^3 \vec{p}_2}{(2 \pi)^3 \sqrt{2 p_2}} \frac{\dd^3 \vec{p}_3}{(2 \pi)^3 \sqrt{2 p_3}} \frac{\dd^3 \vec{p}_4}{(2 \pi)^3 \sqrt{2 p_4}} \times \sqrt{p_1p_2p_3p_4} \, \mathcal{V}(\vec{p}_1,\vec{p}_2,\vec{p}_3,\vec{p}_4) \\ \times (2 \pi)^3 \, \delta^{(3)}(\vec{p}_1 + \vec{p}_2 - \vec{p}_3 - \vec{p}_4)
     \, \had_{\alpha}(\vec{p}_1) 
    \ha_{\alpha}(\vec{p}_3)\had_{\beta}(\vec{p}_2) \ha_{\beta}(\vec{p}_4) \, ,
\end{multline}
with
\begin{equation}
    \mathcal{V}(\vec{p}_1,\vec{p}_2,\vec{p}_3,\vec{p}_4) = \frac{1}{8 \sqrt{p_1 p_2 p_3 p_4}} \left[\bar{u}^{(-)}_{\vec{p}_1}\gamma_{\mu} (1 -\gamma_5) u^{(-)}_{\vec{p}_3}\right] \left[\bar{u}^{(-)}_{\vec{p}_2} \gamma^{\mu} (1 -\gamma_5) u^{(-)}_{\vec{p}_4}\right] \, .
\end{equation}
Finally, we obtain the expression~\eqref{eq:v_nunu_general} after using the relation, derived from Eq.~\eqref{eq:spinors},
\begin{equation}
\label{eq:amplitude}
    \bar{u}^{(-)}_{\vec{p}} \gamma^{\mu} (1 -\gamma_5) u^{(-)}_{\vec{q}} = \frac{2 p q}{\sqrt{(p-p_z)(q-q_z)}} \begin{pmatrix} 
    (1-\hat{p}_z)(1-\hat{q}_z) + (\hat{p}_{x} - \i \hat{p}_y)(\hat{q}_x + \i \hat{q}_y) \\ \\
    (1- \hat{p}_z)(\hat{q}_x + \i \hat{q}_y) + (1- \hat{q}_z)(\hat{p}_x - \i \hat{p}_y) \\ \\
    \i \left[-(1- \hat{p}_z)(\hat{q}_x + \i \hat{q}_y) + (1- \hat{q}_z)(\hat{p}_x - \i \hat{p}_y)\right] \\ \\
    - (1-\hat{p}_z)(1-\hat{q}_z) + (\hat{p}_x - \i \hat{p}_y)(\hat{q}_x + \i \hat{q}_y)
    \end{pmatrix} \, .
\end{equation}
where each row on the right-hand side corresponds to $\mu=0,\dots, 3$.

\subsection{Specific geometries}

For practical purposes, we make explicit the expression of the interaction coefficient $\mathcal{V}$ from Eq.~\eqref{eq:v_nunu_general} in various geometries.

\subsubsection{One dimension}

For a one-dimensional chain of momenta $\vec{p}_k = k \vec{p}_0$, $k \in \{-N/2, \dots , -1, 1, \dots, N/2\}$, we can simplify the general expression~\eqref{eq:v_nunu_general}. Let's take, without loss of generality, $\vec{p}_0 = p_0 \vec{u}_x$. We then have:
\begin{equation}
    \mathcal{V}_\mathrm{1D}(i_1,i_2,i_3,i_4) = \left[\hat{p}_{i_2,x}-\hat{p}_{i_1,x}\right]\left[\hat{p}_{i_4,x}-\hat{p}_{i_3,x}\right] \, . 
\end{equation}
Since, in one dimension, $\hat{p}_{i,x} = \pm 1$, $\mathcal{V}_\mathrm{1D}(i_1,i_2,i_3,i_4)=0$ if $(\vec{p}_{i_1}, \vec{p}_{i_2})$ or $(\vec{p}_{i_3},\vec{p}_{i_4})$ are in the same direction, otherwise $\mathcal{V}_\mathrm{1D}(i_1,i_2,i_3,i_4) = \pm 4$ depending on the respective orientation of $\vec{p}_{i_1}$ and $\vec{p}_{i_3}$.

\subsubsection{Two dimensions}

We consider a uniform grid of momenta in the $(x,y)$ plane. We can then write:
\begin{equation}
    \mathcal{V}_\mathrm{2D}(i_1,i_2,i_3,i_4) = \left[(\hat{p}_{i_2,x} - \i \hat{p}_{i_2,y}) - (\hat{p}_{i_1,x}-\i \hat{p}_{i_1,y})\right] \times \left[(\hat{p}_{i_4,x} + \i \hat{p}_{i_4,y}) - (\hat{p}_{i_3,x}+\i \hat{p}_{i_3,y})\right] \, .
\end{equation}

\subsubsection{Three dimensions}

With three dimensions, we must handle the apparent singularity of Eq.~\eqref{eq:v_nunu_general} when $\vec{p}_j \parallel \vec{u}_z$. To do so, let's use spherical coordinates $\vec{p}_j = p_j \left(\sin \theta_j \cos \phi_j, \sin \theta_j \sin \phi_j, \cos \theta_j\right)$. Then we have:
\begin{align*}
    \hat{p}_{j,x} \pm \i \hat{p}_{j,y} &= \sin(\theta_j) e^{\pm \i \phi_j} = 2 \sin\left(\frac{\theta_j}{2}\right) \cos\left(\frac{\theta_j}{2}\right) e^{\pm \i \phi_j} \, , \\
    1 - \hat{p}_{j,z} &= 1 - \cos(\theta_j) = 2 \sin^2\left(\frac{\theta_j}{2}\right) \, .
\end{align*}
As a consequence,
\[ \lim_{\theta_j \to 0} \left \lvert \frac{\hat{p}_{j,x} \pm \i \hat{p}_{j,y}}{\sqrt{1-\hat{p}_{j,z}}}\right \rvert = \sqrt{2} \, .\]
In other words, the useful limit reads, for instance:
\begin{equation}
\frac{\left[ (1 - \hat{p}_{1,z})(\hat{p}_{2,x}-\i \hat{p}_{2,y}) - (\hat{p}_{1,x} - \i \hat{p}_{1,y})(1 - \hat{p}_{2,z})\right]}{\sqrt{(1-\hat{p}_{1,z})(1-\hat{p}_{2,z})}}
\xrightarrow[\hat{p}_{1,z} \to 1]{} - \sqrt{2} \sqrt{1-\hat{p}_{2,z}} \, .
\end{equation}
We also write the general expression of $\mathcal{V}$ in spherical coordinates,
\begin{multline}
\label{eq:V_spherical}
    \mathcal{V}(\vec{p}_1,\vec{p}_2,\vec{p}_3,\vec{p}_4) = 4 \left[e^{-\i \phi_2}\sin\left(\frac{\theta_1}{2}\right) \cos\left(\frac{\theta_2}{2}\right) - e^{- \i \phi_1} \cos\left(\frac{\theta_1}{2}\right) \sin\left(\frac{\theta_2}{2}\right)\right] \\
    \times \left[e^{\i \phi_4}\sin\left(\frac{\theta_3}{2}\right) \cos\left(\frac{\theta_4}{2}\right) - e^{\i \phi_3} \cos\left(\frac{\theta_3}{2}\right) \sin\left(\frac{\theta_4}{2}\right)\right] \, .
\end{multline}

A comparison with~\cite{Cirigliano:2024pnm} shows that we should have the correspondence $\mathcal{V}(\vec{p}_1,\vec{p}_2,\vec{p}_3,\vec{p}_4) \longleftrightarrow 2 g(\vec{p}_1,\vec{p}_3,\vec{p}_2,\vec{p}_4) = 2 f^\dagger(\vec{p}_1,\vec{p}_2)f(\vec{p}_3,\vec{p}_4)$, where $g$ and $f$ are the functions introduced in~\cite{Cirigliano:2024pnm}. We note that we have an overall phase difference, which is due to a difference in our spinors' normalization: we have $u_{\vec{p}}^{(-)}\big\rvert_{\text{[this work]}} = - \sqrt{2 p} e^{- \i \phi_{\vec{p}}}  u(\vec{p},-)\big\rvert_{\text{\cite{Cirigliano:2024pnm}}}$.

\appsection{Discrete Quantum Kinetic Equation}
\label{app:QKE}

In this Appendix, we introduce the version of the QKE adapted for a fixed grid of momenta, and determine analytically the expected equilibrium state. Compared to the standard collision term with matter in infinite volume which leads to an isotropic Fermi-Dirac distribution, the closed system of interacting neutrinos we consider here is much more constrained, which is reflected in the asymptotic state.

\vspace{-0.3cm}

\subsection{Collision term}

We start from the continuous version of the Quantum Kinetic Equation~\cite{Blaschke:2016xxt,Froustey:2020mcq,Froustey:2022sla}, where the collision term reads, for a gas of ultrarelativistic interacting neutrinos (not including antineutrinos, as in the rest of this paper),
\begin{multline}
    \mathcal{C}(\vec{p}_1) = \frac{1}{2} \frac{2^5 G_F^2}{2 p_1} \int{\frac{\dd^3 \vec{p}_2}{(2 \pi)^3 2 p_2}\frac{\dd^3 \vec{p}_3}{(2 \pi)^3 2 p_3}\frac{\dd^3 \vec{p}_4}{(2 \pi)^3 2 p_4} \, (2 \pi)^3 \delta^{(3)}(\vec{p}_1 + \vec{p}_2 - \vec{p}_3 - \vec{p}_4) \, (2 \pi) \delta(p_1 + p_2 - p_3 - p_4)} \\
    \times p_1 p_2 p_3 p_4 \, (1- \hat{\vec{p}}_1 \cdot \hat{\vec{p}}_2)(1- \hat{\vec{p}}_3\cdot \hat{\vec{p}}_4) \times F_\mathrm{sc}(\nu^{(1)},\nu^{(2)},\nu^{(3)},\nu^{(4)}) \, ,
\end{multline}
with the scattering statistical factor
\begin{equation}
\begin{aligned}
    F_\mathrm{sc}(\nu^{(1)},\nu^{(2)},\nu^{(3)},\nu^{(4)}) &= \Big\{f_4 (\Id-f_2) + \mathrm{Tr}\left[f_4 (\Id-f_2)\right]\Big\} f_3 (\Id - f_1) + (\Id - f_1)f_3 \Big\{(\Id-f_2)f_4 + \mathrm{Tr}\left[(\Id-f_2)f_4\right]\Big\} \\
    &- \Big\{(\Id-f_4) f_2 + \mathrm{Tr}\left[(\Id - f_4) f_2\right]\Big\} (\Id - f_3) f_1 - f_1(\Id - f_3) \Big\{f_2(\Id - f_4) + \mathrm{Tr}\left[f_2(\Id - f_4)\right]\Big\} \, .
\end{aligned}
\end{equation}
We used the shorthand notations $f_i \equiv f(\vec{p}_i)$, and $\Id$ denotes the identity matrix in flavor (or mass) space. Note that, since we do not consider the interaction with charged leptons, the collision term has the exact same form in the mass and flavor bases~\cite{Akita:2020szl}.

Upon discretization on the momentum grid~\eqref{eq:pgrid}, we have the replacement rules~\eqref{eq:discrete_sum} and \eqref{eq:discrete_delta}. Similarly to the latter, the energy-conserving delta function brings an additional factor $L = V^{1/3}$, such that the discrete version of the collision term reads
\begin{equation}
    \label{eq:discrete_collision}
    \mathcal{C}_i = \left(\frac{G_F}{V}\right)^2 V^{1/3} \sum_{i_2} \sum_{\substack{(i_3, i_4) \text{ satisfying}\\ \text{energy-momentum conservation}}}  (1- \hat{\vec{p}}_i \cdot \hat{\vec{p}}_{i_2})(1- \hat{\vec{p}}_{i_3}\cdot \hat{\vec{p}}_{i_4}) \times F_\mathrm{sc}(\nu^{(i)},\nu^{(i_2)},\nu^{(i_3)},\nu^{(i_4)}) \, .
\end{equation}

\subsection{Equilibrium state}

\subsubsection{Classical case}

We first discuss the “classical” evolution of a neutrino gas, that is, where the discrete QKE is reduced to the collision term (discrete Boltzmann equation). In a continuous system where neutrinos can interact with background matter, the equilibrium state corresponds to isotropic Fermi-Dirac distributions. The system we study here is much more constrained, such that the equilibrium state can be determined in a generalized grand-canonical ensemble.

Specifically, the collision term is such that: \emph{(i)} the total kinetic energy, $E_\nu$, is conserved, \emph{(ii)} the total momentum, $\vec{P}_\nu$, is conserved; \emph{(iii)} the individual lepton numbers (so, here, the total number of neutrinos for each flavor, $N_\alpha$) are conserved. We can then find the equilibrium state by maximizing the entropy functional, implementing the constraints via Lagrange multipliers, that is,
\begin{multline}
   \mathsf{S} = \sum_{\alpha,\vec{p}} \left[f_\alpha \ln(f_\alpha) + (1- f_\alpha) \ln(1 - f_\alpha)\right] - \sum_{\alpha} \xi_\alpha \left[\sum_{\vec{p}}{f_\alpha(\vec{p})} - N_\alpha\right]  \\
   + \beta \left[\sum_{\alpha, \vec{p}}{|\vec{p}| \, f_\alpha(\vec{p})} - E_\nu\right] + \vec{\lambda} \cdot \left[\sum_{\alpha, \vec{p}}{\vec{p} \, f_\alpha(\vec{p})} - \vec{P}_\nu\right] \, .
\end{multline}
Maximizing over $f_\alpha(\vec{p})$ leads to the equilibrium distribution
\begin{equation}
    f_\alpha^\mathrm{(eq)}(\vec{p}) = \frac{1}{e^{-\xi_\alpha + \beta |\vec{p}| + \vec{\lambda} \cdot \vec{p}} + 1} \, ,
\end{equation}
where the values of $\{\xi_\alpha, \beta, \vec{\lambda}\}$ are set by the various constraints.

\subsubsection{Including flavor mixing}

If one includes flavor mixing, the equilibrium distributions are better defined in the mass basis. We consider that in the asymptotic state, the density matrix in the mass basis is diagonal (at least, if one averages over the fast vacuum oscillations), and the mass distributions can be determined following the exact same steps, the only difference being that the conserved lepton numbers are now introduced in the mass basis, $N_\alpha \to N_a$. The equilibrium distributions in the flavor basis then read
\begin{equation}
\label{eq:eqb_QKE}
    f_\alpha^\mathrm{(eq)}(\vec{p}) = \sum_{a} \frac{|U_{\alpha a}|^2}{e^{- \xi_a + \beta |\vec{p}| + \vec{\lambda} \cdot \vec{p}}+1} \, .
\end{equation}

\subsubsection{Illustration}

\begin{figure}[!ht]
    \centering
    \includegraphics[width=0.9\textwidth]{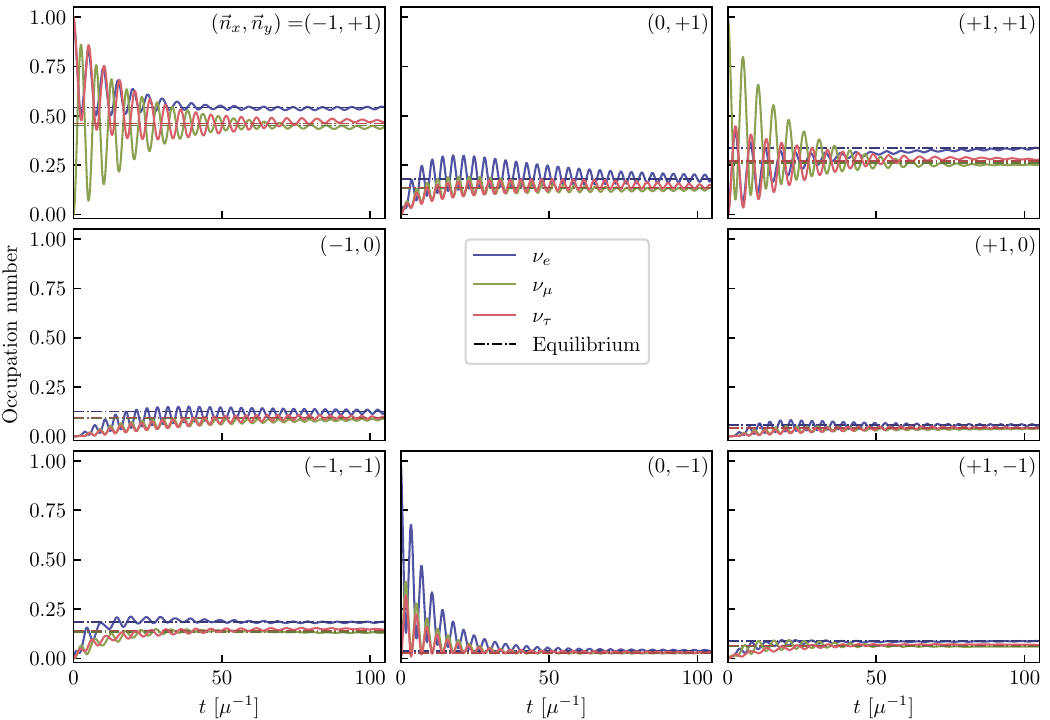}
    \caption{Time evolution of the occupation numbers for a discrete QKE calculation with the collision term multiplied by $\kappa = 3 \times 10^{10}$, starting from the state \eqref{eq:initial_4beams}. The dash-dotted lines are the equilibrium predictions~\eqref{eq:eqb_QKE}, different for each flavor (contrary to Fig.~\ref{fig:3beams_QKE}).}
    \label{fig:4beams_QKE}
\end{figure}

The equilibrium prediction~\eqref{eq:eqb_QKE} is shown in the main text for the 3-neutrino case we study, see Fig.~\ref{fig:3beams_QKE}. However, because of the symmetries of the initial state, the asymptotic values of the occupation numbers are the same for the three flavors. In order to illustrate that this is not always the case, we show in Fig.~\ref{fig:4beams_QKE} the same QKE calculation and equilibrium prediction, but for the initial state
\begin{equation}
    \label{eq:initial_4beams}
|\psi_0\rangle = \hat{\mathcal{A}} \ |\nu_e,(-1,+1) \rangle \otimes |\nu_\tau,(-1,+1) \rangle \otimes |\nu_e,(0,-1)\rangle \otimes |\nu_\mu,(+1,+1) \rangle \, .
\end{equation}
We added an electron neutrino in the $(-1,+1)$ momentum bin compared to \eqref{eq:initial_3beams}.

There is once again a perfect agreement between the theoretical prediction and the asymptotic state reached in this numerical calculation.

\appsection{“2 beams” initial configuration}
\label{app:2beams}

While in the main text we considered a three-neutrino initial configuration, we present here the results for a simpler system. There are two initially occupied states:
\begin{equation}
\label{eq:initial_2beams}
    |\psi_0\rangle = \hat{\mathcal{A}} \, |\nu_e,(-1,-1)\rangle \otimes |\nu_\mu, (+1,+1)\rangle \, .
\end{equation}
This corresponds schematically to two “beams,” one of electron neutrinos in the $(-1,-1)$ direction and one of muon neutrinos in the $(+1,+1)$ direction, see Fig.~\ref{fig:pgrid}.

\begin{figure}[ht]
    \centering
    \begin{minipage}{0.4\linewidth}
        \centering
        \includegraphics[width=\linewidth]{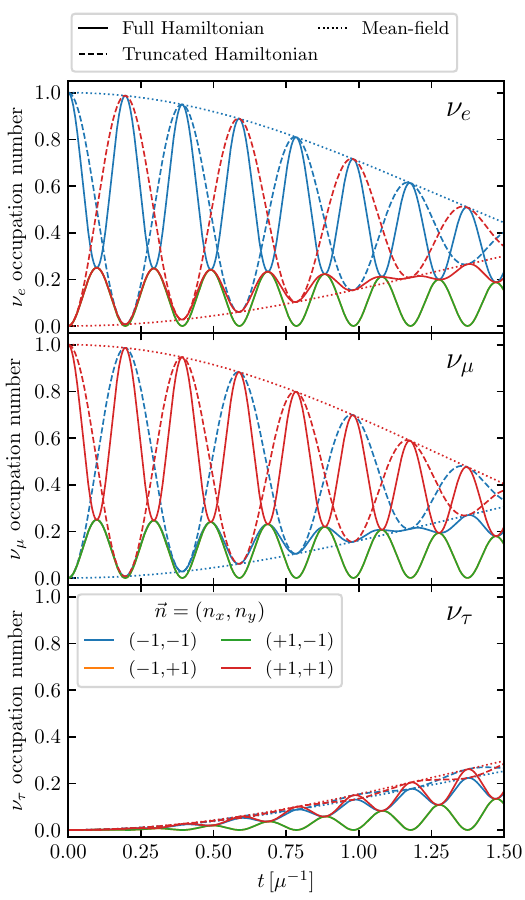}
    \end{minipage}
    \hspace{0.5cm}
    \begin{minipage}{0.45\linewidth}
        \centering
        \includegraphics[width=\linewidth]{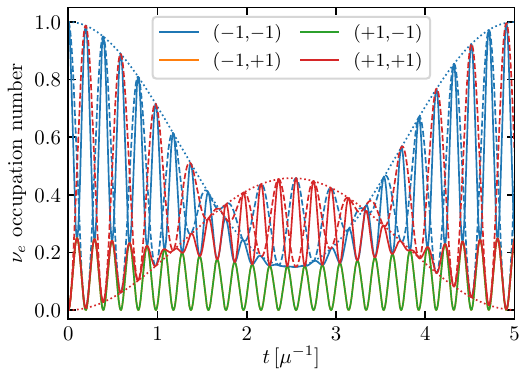}
    \end{minipage}
    \caption{\emph{Left:} Time evolution for the initial state \eqref{eq:initial_2beams}. Many-body calculations using the full (forward/exchange) Hamiltonian are shown with solid (dashed) lines, while the mean-field result is shown with a dotted line. Different colors are used for one-particle states with different momenta. \emph{Right:} Longer time evolution, displaying only the $\nu_e$ occupation numbers.}
    \label{fig:2beams}
\end{figure}

The time evolution of the occupation numbers is shown in Fig.~\ref{fig:2beams}. Like in the main text, we compare three implementations: the mean-field treatment (dotted lines), the many-body calculation with the truncated Hamiltonian (dashed lines), and the full many-body calculation (solid lines).

We can make several observations. First, the mean-field evolution is entirely due to the vacuum term, which is a feature of this very simple geometry (see discussion in Sec.~IV of Ref.~\cite{Rrapaj:2019pxz}). Then, given the symmetry of the initial configuration, the occupation numbers in the $(-1,+1)$ and $(+1,-1)$ bins—which are nonzero in the full many-body case only—have identical evolutions. Finally, there is an acceleration of flavor conversion between the truncated and full many-body calculations (compare the solid and dashed lines at short times), a behavior also observed in~\cite{Cirigliano:2024pnm}.

\newpage

\appsection{Auxiliary results for Trotter error analysis}
\label{app:Trotter_details}

In this Appendix, we present the explicit calculations leading to the estimates of Trotter error bounds described in Sec.~\ref{subsec:Trotter}.

\subsection{One-body/Two-body splitting}
\label{subsec:C12}

\subsubsection{Forward/exchange part}

We detail here the steps leading to the estimate~\eqref{eq:C12_forward}. We start from the explicit expression of $C_{12}^\mathrm{(f/e)}$, defined in Eq.~\eqref{eq:C12_forward_def}:
\begin{equation}
    C_{12}^\mathrm{(f/e)} = 2 \mu \sum_{\alpha, \beta, \gamma} \sum_{i_1 \neq i_2, i}{\omega_{\gamma,i} (1- \hat{\vec{p}}_{i_1}\cdot \hat{\vec{p}}_{i_2})} \left[\had_{\gamma,i} \ha_{\gamma,i} \, , \, \had_{\alpha,i_1} \ha_{\alpha,i_1} \had_{\beta,i_2}\ha_{\beta,i_2} - \had_{\alpha,i_1} \ha_{\alpha, i_2}\had_{\beta,i_2}\ha_{\beta,i_1}\right] \, .
\end{equation}
The first term involves only products of the number operators $\had_{\dots} \ha_{\dots}$, which commute. Only the “exchange” part gives a nonzero contribution. This requires $\gamma = \alpha$ or $\beta$ (and $\alpha \neq \beta$). After some operator algebra, one arrives at the following expression,
\begin{equation}
\left[\had_{\alpha,i} \ha_{\alpha,i}, \had_{\alpha,i_1} \ha_{\alpha, i_2}\had_{\beta,i_2}\ha_{\beta,i_1}\right]  = (\delta_{i i_1} - \delta_{i i_2}) \had_{\alpha,i_1} \ha_{\alpha,i_2} \had_{\beta,i_2} \ha_{\beta,i_1} \, .
\end{equation}
The same procedure applies to $\gamma = \beta$ and leads to an overall minus sign.

Therefore, we have
\begin{equation}
\left[\had_{\gamma,i} \ha_{\gamma,i}, \had_{\alpha,i_1} \ha_{\alpha, i_2}\had_{\beta,i_2}\ha_{\beta,i_1}\right] = (\delta_{\alpha \gamma} - \delta_{\beta \gamma})(\delta_{i i_1} - \delta_{i i_2}) \had_{\alpha,i_1} \ha_{\alpha,i_2} \had_{\beta,i_2} \ha_{\beta,i_1} \, ,
\end{equation}
such that (we get a factor two by rewriting half the terms through the simultaneous changes $\alpha \leftrightarrow \beta$ and $i_1 \leftrightarrow i_2$)
\begin{equation}
    C_{12}^\mathrm{(f/e)} = 4 \mu \sum_{\alpha \neq \beta} \sum_{i_1 \neq i_2}{(1- \hat{\vec{p}}_{i_1}\cdot \hat{\vec{p}}_{i_2}) (\omega_{\alpha i_1} - \omega_{\alpha i_2})} \had_{\alpha,i_1} \ha_{\alpha, i_2} \had_{\beta,i_2} \ha_{\beta,i_1} \, .
\end{equation}
The flavor sum contains $N_F(N_F-1)$ terms, and the momentum one $N(N-1)$. Using $|1- \hat{\vec{p}}_{i_1}\cdot \hat{\vec{p}}_{i_2}| \leq 2$ and $\lVert \had \cdots \ha \rVert \leq 1$, we then get the bound~\eqref{eq:C12_forward}.

\subsubsection{Non-forward part}

We now focus on the non-forward/exchange parts of the two-body Hamiltonian, leading to the estimate~\eqref{eq:C12_nonforward}. The associated Trotter error reads
\begin{equation}
    C_{12}^\stkout{\mathrm{(f/e)}} = \mu \sum_{\alpha, \beta, \gamma} \sum_{i, i_1 \neq i_2} \sum_{i_3 \notin \{i_1, i_2\}}{\omega_{\gamma,i} \, \mathcal{V}(\vec{p}_{i_1},\vec{p}_{i_2},\vec{p}_{i_3},\vec{p}_{i_4})} \delta_{\vec{p}_{i_4} = \vec{p}_{i_1}+\vec{p}_{i_2}-\vec{p}_{i_3}} \left[\had_{\gamma,i} \ha_{\gamma,i} \, , \, \had_{\alpha,i_1} \ha_{\alpha,i_3} \had_{\beta,i_2}\ha_{\beta,i_4}\right] \, .
\end{equation}
If $\alpha \neq \beta$,
\begin{equation}
\left[\had_{\alpha,i} \ha_{\alpha,i} \, , \, \had_{\alpha,i_1} \ha_{\alpha, i_3}\had_{\beta,i_2}\ha_{\beta,i_4}\right] = (\delta_{i i_1} - \delta_{i i_3}) \had_{\alpha,i_1} \ha_{\alpha,i_3} \had_{\beta,i_2} \ha_{\beta,i_4} \, ,
\end{equation}
such that (and the following equation is also true if $\alpha = \beta$)
\begin{equation}
\left[\had_{\gamma,i} \ha_{\gamma,i} \, , \, \had_{\alpha,i_1} \ha_{\alpha, i_3}\had_{\beta,i_2}\ha_{\beta,i_4}\right] = \left[\delta_{\alpha \gamma}(\delta_{i i_1} - \delta_{i i_3}) + \delta_{\beta \gamma} (\delta_{i i_2} - \delta_{i i_4}) \right] \had_{\alpha,i_1} \ha_{\alpha,i_3} \had_{\beta,i_2} \ha_{\beta,i_4} \, .
\end{equation}
Therefore, we get
\begin{equation}
    \label{eq:C12_intermediate}
    C_{12}^\stkout{\mathrm{(f/e)}} = \mu \sum_{\alpha, \beta} \sum_{i_1 \neq i_2} \sum_{i_3 \notin \{i_1, i_2\}}{\mathcal{V}(\vec{p}_{i_1},\vec{p}_{i_2},\vec{p}_{i_3},\vec{p}_{i_4}) \left(\omega_{\alpha i_1} - \omega_{\alpha i_3} + \omega_{\beta i_2} - \omega_{\beta i_4}\right)} \, \delta_{\vec{p}_{i_4} = \vec{p}_{i_1}+\vec{p}_{i_2}-\vec{p}_{i_3}}  \, \had_{\alpha,i_1} \ha_{\alpha,i_3} \had_{\beta,i_2} \ha_{\beta,i_4} \, .
\end{equation}
A crude bound can be obtained by using $|\mathcal{V}| \leq 16$, a simple consequence of the expression~\eqref{eq:V_spherical}. The flavor sums give a factor $N_F^2$, the sums over $i_1$ and $i_2$ give a factor $N(N-1)$, and the last sum contains $N_\mathrm{PS}$ terms. This ``phase-space number'' is a bound on the number of pairs $(i_3,i_4)$ which, for a given $(i_1,i_2)$, satisfy energy and momentum conservation. For the limited grid we consider in Fig.~\ref{fig:pgrid} or even the ones used for the scaling studies in Fig.~\ref{fig:quantum_resources}, this number is actually of order unity. Indeed, one needs a very dense momentum grid to have more than one or two momentum pairs which have the same total kinetic energy and momentum. If one goes to the large-$N$ limit, it is possible to estimate $N_\mathrm{PS}$ in the continuum limit, see Appendix~\ref{app:phase_space}.

\subsection{Two-body splitting}
\label{app:commutator}

The Trotter error associated with the splitting of $H_{\nu \nu}$ into individual components is given in Eq.~\eqref{eq:C22_def}, which we recall here in its full form:
\begin{equation}
    \label{eq:C22_app}
   \lVert C_{22} \rVert = \frac12 \mu^2 \sum_{\alpha, \beta, \gamma,\sigma} \sum_{i_1 \neq i_2, j_1 \neq j_2} \sum_{i_3, j_3} |\mathcal{V}(\vec{p}_{i_1},\vec{p}_{i_2},\vec{p}_{i_3},\vec{p}_{i_4}) \mathcal{V}(\vec{p}_{j_1},\vec{p}_{j_2},\vec{p}_{j_3},\vec{p}_{j_4})|  \left\lVert\left[\had_{\alpha,i_1} \ha_{\alpha,i_3} \had_{\beta,i_2} \ha_{\beta,i_4}, \had_{\gamma,j_1} \ha_{\gamma,j_3} \had_{\sigma,j_2} \ha_{\sigma,j_4}\right] \right\rVert \, .
\end{equation}
Several anticommutations of $\ha$, $\had$ operators allow one to rewrite
\begin{equation}
\label{eq:commutator_general}
\begin{aligned}
    \left[\had_{\alpha,i_1} \ha_{\alpha,i_3} \had_{\beta,i_2} \ha_{\beta,i_4}, \had_{\gamma,j_1} \ha_{\gamma,j_3} \had_{\sigma,j_2} \ha_{\sigma,j_4}\right] &= \delta_{\beta \gamma} \delta_{i_4 j_1} \,  \had_{\alpha,i_1} \ha_{\alpha,i_3} \had_{\beta,i_2} \ha_{\gamma,j_3} \had_{\sigma,j_2} \ha_{\sigma,j_4} \\
    &+ \delta_{\alpha \gamma} \delta_{i_3 j_1} \, \had_{\alpha,i_1} \had_{\beta,i_2} \ha_{\beta,i_4} \ha_{\gamma,j_3} \had_{\sigma,j_2} \ha_{\sigma,j_4} \\
    &- \delta_{\beta \gamma} \delta_{i_2 j_3} \, \had_{\gamma,j_1} \had_{\alpha,i_1} \ha_{\alpha,i_3} \ha_{\beta,i_4} \had_{\sigma,j_2} \ha_{\sigma,j_4} \\
    &- \delta_{\alpha \gamma} \delta_{i_1 j_3} \, \had_{\gamma,j_1} \ha_{\alpha,i_3} \had_{\beta,i_2} \ha_{\beta,i_4} \had_{\sigma,j_2} \ha_{\sigma,j_4} \\
    &+ \delta_{\beta \sigma} \delta_{i_4 j_2} \, \had_{\gamma,j_1} \ha_{\gamma,j_3} \had_{\alpha,i_1} \ha_{\alpha,i_3} \had_{\beta,i_2} \ha_{\sigma,j_4} \\
    &+ \delta_{\alpha \sigma} \delta_{i_3 j_2} \, \had_{\gamma,j_1} \ha_{\gamma,j_3} \had_{\alpha,i_1} \had_{\beta,i_2} \ha_{\beta,i_4} \ha_{\sigma,j_4}  \\
    &- \delta_{\beta \sigma} \delta_{i_2 j_4} \, \had_{\gamma,j_1} \ha_{\gamma,j_3}  \had_{\sigma,j_2} \had_{\alpha,i_1} \ha_{\alpha,i_3} \ha_{\beta,i_4} \\
    &- \delta_{\alpha \sigma} \delta_{i_1 j_4} \, \had_{\gamma,j_1} \ha_{\gamma,j_3} \had_{\sigma,j_2} \ha_{\alpha,i_3} \had_{\beta,i_2} \ha_{\beta,i_4} \, .
\end{aligned}
\end{equation}
By going through painstaking case separation, we could identify in which cases this commutator is zero. We only want a conservative upper bound, which we can obtain by bounding each of the 8 contributions on the right-hand side by 1, and noting that the Kronecker deltas remove one flavor sum and one momentum sum.

A bound on~\eqref{eq:C22_app} can thus be obtained as follows: a factor $N_F^3$ for the flavor sums [since one is removed by Eq.~\eqref{eq:commutator_general}], a factor $N^3$ for the momentum sums, and a factor $N_\mathrm{PS}^2$ for the sum over $i_3,j_3$ [see the discussion after Eq.~\eqref{eq:C12_intermediate}]. Finally, the product of $\mathcal{V}$ can be bounded by $16^2$, which leads to
\begin{equation}
\label{eq:C22_bound}
\lVert C_{22} \rVert \leq \frac12 \mu^2 \times N_F^3 \times N^3 \times N_\mathrm{PS}^2 \times 16^2 \times 8 = 1024 \, \mu^2 \, N_F^3 \, N^3 \, N_\mathrm{PS}^2 \, .
\end{equation}

In the forward/exchange case, where the Hamiltonian is limited to \eqref{eq:H_forward}, we make the following observations:
\begin{itemize}
    \item the commutator of two forward terms is zero, since number operators $\had_i \ha_i$ commute with one another;
    \item the flavors must be different ($\alpha \neq \beta$ and $\gamma \neq \sigma$), otherwise there is no distinction between forward and exchange;
    \item we then once again remove one flavor sum and one momentum sum through the commutator [special case of Eq.~\eqref{eq:commutator_general}].
\end{itemize}
The bound on $C_{22}$ is then similar, but without the factor $N_\mathrm{PS}^2$, and one can use the fact that for forward/exchange terms $|\mathcal{V}| \leq 4$, see Eq.~\eqref{eq:V_forward}.

\subsection{Phase-space integrals}
\label{app:phase_space}

In order to determine the scaling of the Trotter error bounds for large $N$, we need to obtain expressions for the number of modes in the forward and general cases. To that end, we assume here that we are in the continuum limit and that the distribution of momentum modes is isotropic, in a sphere of maximum momentum $|\vec{p}_\mathrm{max}| \equiv \Lambda = (2 \pi /L) \, n_\mathrm{max}$.
We can thus obtain the number of momentum combinations, up to a normalization constant, as phase-space integrals. 

In order to manipulate dimensionless quantities, we use the representation of momenta via the integer vector $\vec{n}$, i.e., $\vec{p} = (2 \pi/L) \vec{n}$, with $\vec{n}$ promoted to a continuous quantity in the large-$N$ limit.

\paragraph*{Forward case.} The number of momentum combinations satisfying spatial momentum conservation ($\vec{p}_1 + \vec{p}_2 = \vec{p}_3 + \vec{p}_4$) and forward kinematics ($\vec{p}_1 = \vec{p}_3$) is
\begin{equation}
\begin{aligned}
    N_\mathrm{forward} &= \int{\dd^3{\vec{n}_1} \, \dd^3{\vec{n}_2} \, \dd^3{\vec{n}_3} \, \dd^3{\vec{n}_4} \, \delta^{(3)}(\vec{n}_1 + \vec{n}_2 - \vec{n}_3 - \vec{n}_4) \, \delta^{(3)}(\vec{n}_3 - \vec{n}_1)} \\
    &= \int{\dd^3{\vec{n}_1} \, \dd^3{\vec{n}_2}} \\
    &= (4 \pi)^2 \int_{0}^{n_\mathrm{max}}{n_1^2 \dd{n_1}} \int_{0}^{n_\mathrm{max}}{n_2^2 \dd{n_2}} \\
    &= \left(\frac43 \pi n_\mathrm{max}^3\right)^2 \, .
\end{aligned}
\end{equation}

\paragraph*{General case, with kinetic energy conservation.} To compute the general number of terms which satisfy energy-momentum conservation, we use a trick from~\cite{Dolgov:1997mb} to rewrite the momentum delta-function by introducing a new variable.

\begin{equation}
    \begin{aligned}
        N_\text{general} &= 
\int{\dd^3{\vec{n}_1} \, \dd^3{\vec{n}_2} \, \dd^3{\vec{n}_3} \, \dd^3{\vec{n}_4} \, \delta(|\vec{n}_1| + |\vec{n}_2| - |\vec{n}_3| - |\vec{n}_4|) \, \delta^{(3)}(\vec{n}_1 + \vec{n}_2 - \vec{n}_3 - \vec{n}_4)} \\
    &= \int{\dd^3{\vec{n}_1} \, \dd^3{\vec{n}_2} \, \dd^3{\vec{n}_3} \, \dd^3{\vec{n}_4} \, \delta(n_1+n_2-n_3-n_4)} \times \int_{\mathbb{R}^3}{\frac{\dd^3{\vec{\lambda}}}{(2 \pi)^3} \, e^{i \vec{\lambda} \cdot (\vec{n}_1 + \vec{n}_2 - \vec{n}_3 - \vec{n}_4)}} \\
    &= \int{n_1^2 \dd{n_1} \, n_2^2 \dd{n_2} \, n_3^2 \dd{n_3} \, n_4^2 \dd{n_4} \, \delta(n_1 + n_2 - n_3 - n_4)} \int_{\mathbb{R}^3}{\frac{\dd^3{\vec{\lambda}}}{(2 \pi)^3}} \left(\int{\dd{\Omega_1} e^{i \vec{\lambda} \cdot \vec{n}_1}}\right) \cdots \left(\int{\dd{\Omega_4} e^{-i \vec{\lambda}\cdot \vec{n}_4}}\right) \, .
    \end{aligned}
\end{equation}
We define locally the spherical coordinates of $\vec{n}_i$ around $\vec{\lambda}$, such that $\vec{\lambda} \cdot \vec{n}_i = \lambda n_i \cos(\theta_i)$, and we use
\begin{equation}
    \int_{0}^{\pi}{\sin(\theta) \, \dd{\theta} \, e^{\pm i \lambda q \cos(\theta)}} = \frac{2}{\lambda q} \sin(\lambda q) \, .
\end{equation}
We thus have
\begin{equation}
    \begin{aligned}
        N_\text{general} &= \int_{0}^{n_\mathrm{max}}{\prod_{i=1}^{4} n_i^2 \dd{n_i} \, \delta(n_1 + n_2 - n_3 - n_4)} \int_{\mathbb{R}^3}{\frac{\dd^3{\vec{\lambda}}}{(2 \pi)^3}} \frac{ (4\pi)^4}{\lambda^4 n_1 n_2 n_3 n_4} \sin(\lambda n_1) \sin(\lambda n_2) \sin(\lambda n_3) \sin(\lambda n_4) \\
        &= (2 \pi)^3 \int_{0}^{n_\mathrm{max}}{\dd{n_1}} \int_{0}^{n_\mathrm{max}}{\dd{n_2}} \int_{0}^{\mathrm{min}(n_1 + n_2, n_\mathrm{max})}{\dd{n_3}} \, n_1 n_2 n_3 (n_1+n_2-n_3) \\
        &\qquad \qquad \qquad \qquad \qquad \times \frac{16}{\pi} \int_{0}^{\infty}{\frac{\dd{\lambda}}{\lambda^2} \sin(\lambda n_1) \sin(\lambda n_2) \sin(\lambda n_3) \sin(\lambda [n_1+n_2-n_3])} \\
        &= (2 \pi)^3 \, \frac{23 n_\mathrm{max}^8}{144} \, ,
    \end{aligned}
\end{equation}
where we get the last result in \texttt{Mathematica}~\cite{Mathematica}. The ratio between the two number of modes then reads
\begin{equation}
    \frac{N_\text{general}}{N_\text{forward}} = \frac{23}{32} \pi \, n_\mathrm{max}^2 \, .
\end{equation}
This ratio is precisely the multiplicative factor transforming a sum over $(i_1,i_2)$ only (forward case) to the sum over $(i_1,i_2,i_3,i_4)$ satisfying energy-momentum conservation, which we called $N_\mathrm{PS}$. Since by definition the number of momentum states is $N = (4/3) \pi n_\mathrm{max}^3$, we have the scaling in the continuum limit:
\begin{equation}
\label{eq:scaling_NPS}
    N_\mathrm{PS} = \frac{N_\text{general}}{N_\text{forward}} \simeq 0.87 \, N^{2/3} \, .
\end{equation}

%
%
%

\end{document}